\documentclass[prc,twocolumn,epsfig,nofootinbib,floatfix,superscriptaddress]{revtex4}
\usepackage{graphics}
\usepackage{epsfig}
\usepackage{amsfonts}
\usepackage{amsmath}
\usepackage{mathrsfs}
\usepackage{float}
\usepackage{bm}
\usepackage{color}
\usepackage{scrextend}
\usepackage{tablefootnote}
\usepackage[breaklinks=true,colorlinks=true,linkcolor=blue,urlcolor=blue,citecolor=red]{hyperref}

\begin{document}
	\title{Isoscalar giant monopole resonance in $^{24}$Mg and $^{28}$Si: Effect of coupling between the isoscalar monopole and quadrupole strength}
	\author{A. Bahini}
	\email[]{armand.bahini@wits.ac.za}
	\affiliation{School of Physics, University of the Witwatersrand, Johannesburg 2050, South Africa}
	\affiliation{iThemba Laboratory for Accelerator Based Sciences, Somerset West 7129, South Africa}
	
	\author{V. O. Nesterenko}
	\email[]{nester@theor.jinr.ru}
	\affiliation{Laboratory of Theoretical Physics, Joint Institute for Nuclear Research, Dubna, Moscow Region 141980, Russia}
	\affiliation{State University ``Dubna'', Dubna, Moscow Region 141980, Russia}
	\affiliation{Moscow Institute of Physics and Technology, Dolgoprudny, Moscow Region 141701, Russia}
	\author{I. T. Usman}
	\affiliation{School of Physics, University of the Witwatersrand, Johannesburg 2050, South Africa}
	\author{P. von Neumann-Cosel}
	\affiliation{Institut f\"{u}r Kernphysik, Technische Universit\"{a}t Darmstadt, D-64289 Darmstadt, Germany}
	\author{R. Neveling}
	\affiliation{iThemba Laboratory for Accelerator Based Sciences, Somerset West 7129, South Africa}
	\author{J. Carter}
	\affiliation{School of Physics, University of the Witwatersrand, Johannesburg 2050, South Africa}
    \author{J. Kvasil}
    \affiliation{Institute of Particle and Nuclear Physics, Charles University, CZ-18000, Prague 8, Czech Republic}
    \author{A. Repko}
    \affiliation{Institute of Particle and Nuclear Physics, Charles University, CZ-18000, Prague 8, Czech Republic}
    \affiliation{Institute of Physics, Slovak Academy of Sciences, 84511, Bratislava, Slovakia}
	\author{P. Adsley}
	\affiliation{School of Physics, University of the Witwatersrand, Johannesburg 2050, South Africa}
	\affiliation{iThemba Laboratory for Accelerator Based Sciences, Somerset West 7129, South Africa}
	\affiliation{\hbox{Department of Physics, Stellenbosch University, 7602 Matieland, Stellenbosch, South Africa}}
	\affiliation{\hbox{ Irene Joliot Curie Lab, UMR8608, IN2P3-CNRS, Universit\'{e} Paris Sud 11, 91406 Orsay, France}}
	\author{N. Botha}
	\affiliation{School of Physics, University of the Witwatersrand, Johannesburg 2050, South Africa}
	\author{J. W. Brummer}
		\affiliation{iThemba Laboratory for Accelerator Based Sciences, Somerset West 7129, South Africa}
	\affiliation{\hbox{Department of Physics, Stellenbosch University, 7602 Matieland, Stellenbosch, South Africa}}
	\author{L. M. Donaldson}
	\affiliation{iThemba Laboratory for Accelerator Based Sciences, Somerset West 7129, South Africa}
		\author{S. Jongile}
		\affiliation{iThemba Laboratory for Accelerator Based Sciences, Somerset West 7129, South Africa}
		\affiliation{\hbox{Department of Physics, Stellenbosch University, 7602 Matieland, Stellenbosch, South Africa}}
	\author{T. C. Khumalo}
	\affiliation{School of Physics, University of the Witwatersrand, Johannesburg 2050, South Africa}
		\affiliation{iThemba Laboratory for Accelerator Based Sciences, Somerset West 7129, South Africa}	
\affiliation{\hbox{Department of Physics, University of Zululand, Richards Bay, 3900, South Africa}}			
	\author{M. B. Latif}	
	\affiliation{School of Physics, University of the Witwatersrand, Johannesburg 2050, South Africa}
	\affiliation{iThemba Laboratory for Accelerator Based Sciences, Somerset West 7129, South Africa}
	\author{K. C. W. Li}
	\affiliation{iThemba Laboratory for Accelerator Based Sciences, Somerset West 7129, South Africa}	
	\affiliation{\hbox{Department of Physics, Stellenbosch University, 7602 Matieland, Stellenbosch, South Africa}}
	\author{P. Z. Mabika}
	\affiliation{Department of Physics and Astronomy, University of the Western Cape, Bellville 7535, South Africa}
\author{P. T. Molema}	
\affiliation{School of Physics, University of the Witwatersrand, Johannesburg 2050, South Africa}
\affiliation{iThemba Laboratory for Accelerator Based Sciences, Somerset West 7129, South Africa}	
	\author{C. S. Moodley}
	\affiliation{School of Physics, University of the Witwatersrand, Johannesburg 2050, South Africa}
		\affiliation{iThemba Laboratory for Accelerator Based Sciences, Somerset West 7129, South Africa}
	\author{S. D. Olorunfunmi}
	\affiliation{School of Physics, University of the Witwatersrand, Johannesburg 2050, South Africa}
		\affiliation{iThemba Laboratory for Accelerator Based Sciences, Somerset West 7129, South Africa}
	\author{P. Papka}
	\affiliation{iThemba Laboratory for Accelerator Based Sciences, Somerset West 7129, South Africa}
	\affiliation{\hbox{Department of Physics, Stellenbosch University, 7602 Matieland, Stellenbosch, South Africa}}
	\author{L. Pellegri}
	\affiliation{School of Physics, University of the Witwatersrand, Johannesburg 2050, South Africa}
	\affiliation{iThemba Laboratory for Accelerator Based Sciences, Somerset West 7129, South Africa}
	\author{B. Rebeiro}
	\affiliation{Department of Physics and Astronomy, University of the Western Cape, Bellville 7535, South Africa}
	\author{ E. Sideras-Haddad}
	\affiliation{School of Physics, University of the Witwatersrand, Johannesburg 2050, South Africa}
	\author{F. D. Smit}
	\affiliation{iThemba Laboratory for Accelerator Based Sciences, Somerset West 7129, South Africa}
	\author{S. Triambak}
	\affiliation{Department of Physics and Astronomy, University of the Western Cape, Bellville 7535, South Africa}
	\author{J. J. van Zyl}
	\affiliation{\hbox{Department of Physics, Stellenbosch University, 7602 Matieland, Stellenbosch, South Africa}}
	\date{\today}
	\begin{abstract}
\noindent \textbf{Background:} In highly deformed nuclei, there is a noticeable coupling of the Isoscalar Giant Monopole Resonance (ISGMR) and the $K = 0$ component of the Isoscalar Giant Quadrupole Resonance (ISGQR), which results in a double peak structure of the isoscalar monopole (IS0) strength (a narrow low-energy deformation-induced peak and a main broad ISGMR part). The energy of the narrow low-lying IS0 peak  is sensitive to both the incompressibility modulus  $K_\infty$ and the coupling between IS0 and isoscalar quadrupole (IS2) strength.

\noindent\textbf{Objective:}  This study aims to investigate the two-peaked structure of the ISGMR in
the prolate $^{24}$Mg and oblate $^{28}$Si nuclei and identify among a variety of energy density functionals based on Skyrme parameterisations the one which best describes the experimental data. This will allow for conclusions regarding the nuclear incompressibility.
Because of the strong IS0/IS2 coupling, the deformation splitting of the ISGQR will also be analysed.

\noindent\textbf{Methods:} The ISGMR was excited in $^{24}$Mg and $^{28}$Si using $\alpha$-particle inelastic scattering measurements acquired with an $E_\alpha = 196$ MeV beam at scattering angles
$\theta_{\text{Lab}} = 0^\circ$ and $4^\circ$.
The K$600$ magnetic spectrometer at iThemba LABS was used to detect and momentum analyse the inelastically scattered $\alpha$ particles. An experimental energy resolution of $\approx 70$ keV (FWHM) was attained, revealing fine structure in the excitation-energy region of the ISGMR. The IS0 strength distributions in the nuclei studied were obtained with the Difference-of-Spectrum (DoS) technique. The theoretical comparison is based on the quasiparticle random-phase approximation (QRPA) with a representative set of Skyrme forces.

\noindent\textbf{Results:} IS0 strength distributions for $^{24}$Mg and $^{28}$Si are extracted and compared to previously published results from experiments with a lower energy resolution. With some exceptions, a reasonable agreement is obtained. The IS0 strength is found to be separated into a narrow structure at about $13-14$ MeV in $^{24}$Mg, $17-19$ MeV in $^{28}$Si and a broad structure at $19-26$ MeV in both nuclei. The data are compared with QRPA results. The results of the calculated characteristics of IS0 states demonstrate the strong IS0/IS2 coupling in strongly prolate $^{24}$Mg and oblate $^{28}$Si. The narrow IS0 peaks are shown to arise due to the deformation-induced IS0/IS2 coupling and strong collective effects. The cluster features of the narrow IS0 peak at $13.87$ MeV in $^{24}$Mg  are also discussed. The best description of the IS0 data is obtained using the Skyrme force SkP$^\delta$ with an associated low nuclear incompressibility $K_\infty = 202$ MeV allowing for  both the energy of the peak and integral IS0 strength  in $^{24}$Mg and $^{28}$Si to be reproduced. The features of the ISGQR in these nuclei are also investigated. An anomalous deformation splitting of the ISGQR in oblate $^{28}$Si is found.

\noindent\textbf{Conclusions:} The ISGMR and ISGQR in light deformed nuclei are coupled and thus need to be 
described simultaneously. Only such a description is relevant and consistent. The deformation-induced narrow IS0 peaks can serve as an additional
sensitive measure of the nuclear incompressibility.
\end{abstract}
	
	\pacs{21.60.Jz, 27.30.+t, 21.10.-k }
	\maketitle
\section{Introduction}
\label{s1}
For many years, the Isoscalar Giant Monopole Resonance (ISGMR) served as an important source
of information on the nuclear incompressibility \cite{Blaizot,Harakeh,GC_review2018}.
In deformed nuclei, this resonance demonstrates a remarkable coupling of  the isoscalar monopole
and quadrupole $(K = 0)$ modes, which leads to a double peak structure of the IS0 strength \cite{abgrall1980,jang1983,nishizaki1985,buenerd1980,garg1984,E0_Kvasil_PRC2016,E0_Colo_PLB2020}
(here, $K$ stands for the projection of the total angular momentum on the symmetry axis of an axially deformed nucleus \cite{bohr1998nuclear}). Strongly deformed light nuclei, like prolate $^{24}$Mg and oblate $^{28}$Si, are especially attractive for an exploration of the nuclear incompressibility $K_\infty$ and the role of the IS0/IS2 coupling. Indeed, in these nuclei, one may expect a deformation-induced strong narrow peak in the IS0 strength located below the main ISGMR region
\cite{24Mg_GG_PLB2015,24Mg_GG_PRC2016,24Mg_PPJC2016}. The energy of this peak should be sensitive
to both the incompressibility $K_{\infty}$ and the IS0/IS2 coupling strength.
	
The ISGMR in $^{24}$Mg has been explored in $(\alpha,\alpha^\prime)$ and ($^6$Li,$^6$Li$^\prime$)
experiments since the $1980$s (see Ref.~\cite{Harakeh}  for the extensive review and discussion). More recent $(\alpha,\alpha^\prime)$ experiments were performed at the Research Center for Nuclear Physics (RCNP) for $^{24}$Mg \cite{kaw2013,24Mg_GG_PLB2015,24Mg_GG_PRC2016} and $^{28}$Si \cite{28Si_PGG_PRC2016}. In $2021$, the results of an RCNP ($^6$Li,$^6$Li$^\prime$) experiment on IS0 strength in $^{24}$Mg were published \cite{Zabora2021}.
The IS0/IS2 coupling was mentioned in Refs.
\cite{24Mg_GG_PLB2015,Zabora2021} in order to explain the existence of the IS0 structure located around $16 - 19$ MeV excitation in $^{24}$Mg. The existence of a prominent IS0 peak at about $18$ MeV in oblate $^{28}$Si was noted but not yet explained \cite{28Si_PGG_PRC2016}. These results generally confirm previous findings from $(\alpha,\alpha^\prime)$ experiments performed by the Texas A\&M University (TAMU)
group on $^{24}$Mg \cite{youngblood2009isoscalar} (for the dataset where the out-of-plane angle was measured) and $^{28}$Si \cite{youngblood2007isoscalar}.
For $^{28}$Si, discrepancies between RCNP \cite{28Si_PGG_PRC2016} and TAMU
\cite{youngblood2007isoscalar} results in the high excitation-energy region above
$20$ MeV were noted and attributed to the phenomenological background subtraction method employed by TAMU \cite{GC_review2018}.
	
The IS0/IS2 coupling  was earlier considered within the quasiparticle random phase approximation (QRPA) in $^{24}$Mg \cite{24Mg_GG_PLB2015,24Mg_GG_PRC2016,24Mg_PPJC2016,Zabora2021,yoshida2010,adsley2021isoscalar}
and $^{28}$Si \cite{adsley2021isoscalar}. However,  in the study  \cite{adsley2021isoscalar}, the narrow IS0 peaks in $^{24}$Mg and $^{28}$Si were not specifically discussed. Furthermore, one can dispute whether the low-energy narrow IS0 peak in $^{24}$Mg has been correctly assigned (and used in the QRPA analysis) in the RCNP studies
\cite{24Mg_GG_PLB2015,24Mg_GG_PRC2016,Zabora2021}
(see discussion in Sect. \ref{s42}).

In this paper, the ISGMR in prolate $^{24}$Mg and oblate $^{28}$Si is explored using $\alpha$-particle inelastic scattering measurements with an $E_\alpha = 196$ MeV beam measured at scattering angles $\theta_{\text{Lab}} = 0^\circ$ and $4^\circ$
using the high energy-resolution K$600$ magnetic spectrometer at the iThemba Laboratory for Accelerator Based Science (LABS), South Africa. The study focuses on the deformation-induced splitting of the ISGMR into two parts: a narrow peak arising due to IS0/IS2 coupling, and a wide structure representing the main ISGMR. The former exists only in deformed nuclei while the latter appears in both spherical and deformed nuclei.

By comparing the experimental data obtained with QRPA calculations, we will identify the narrow IS0 peak in $^{24}$Mg, explain the origin of the narrow IS0 peaks in both  prolate $^{24}$Mg and oblate $^{28}$Si, and suggest a Skyrme force suitable for the description  of IS0 strength in these nuclei. Moreover, we will analyse in detail the IS2 strength, which determines the energy of the narrow IS0 peak, and show that, in strongly deformed nuclei,  only a simultaneous description of IS0 and IS2 strengths can be considered as consistent. The simultaneous availability of data for a prolate ($^{24}$Mg) and an oblate ($^{28}$Si) case allows for the generalisation of the conclusions.
	
The paper is organised as follows: in Sec. \ref{s2}, the experimental details are given. The method of extraction and the resulting IS0 strength distributions are presented in Sec. \ref{s3} while, in Sec. \ref{s4}, details of the QRPA calculations are given and the comparison with the data is shown and discussed. In Sec. \ref{s5}, conclusions are drawn and finally in Appendix \ref{s6}, the predicted deformation splitting of the ISGQR in $^{24}$Mg and $^{28}$Si is presented.

	\begin{figure*}[t] 
		\centering
		\includegraphics[width=13cm]{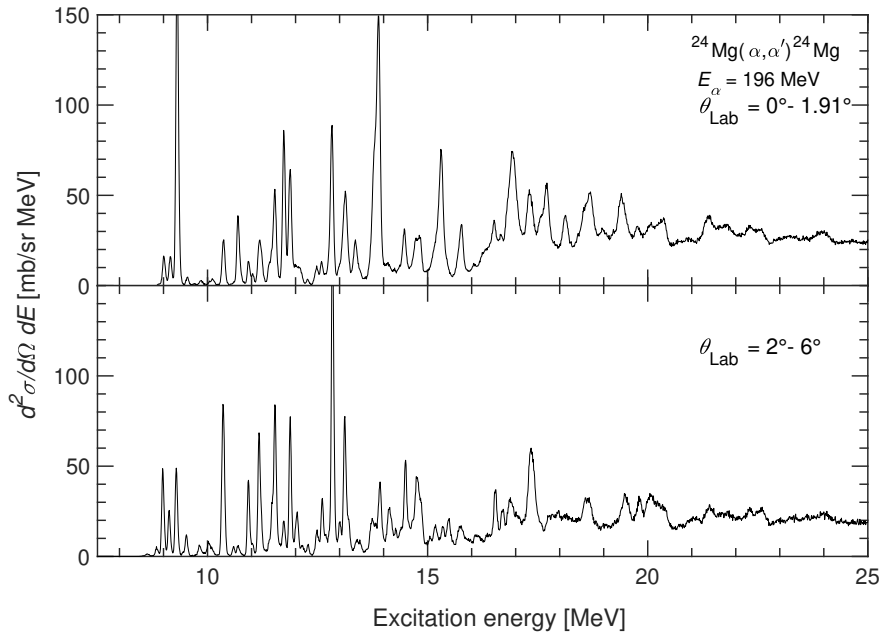}
		\caption{Double-differential cross sections for the $^{24}$Mg($\alpha,\alpha^\prime$) reaction at $E_\alpha = 196$ MeV with $\theta_{\text{Lab}} = 0^{\circ} - 1.91^{\circ}$ (top) and $\theta_{\text{Lab}} = 2^{\circ} - 6^{\circ}$ (bottom).}
		\label{FIG:1a}
	\end{figure*}
	
	\begin{figure*}[t] 
		\centering
		\includegraphics[width=13cm]{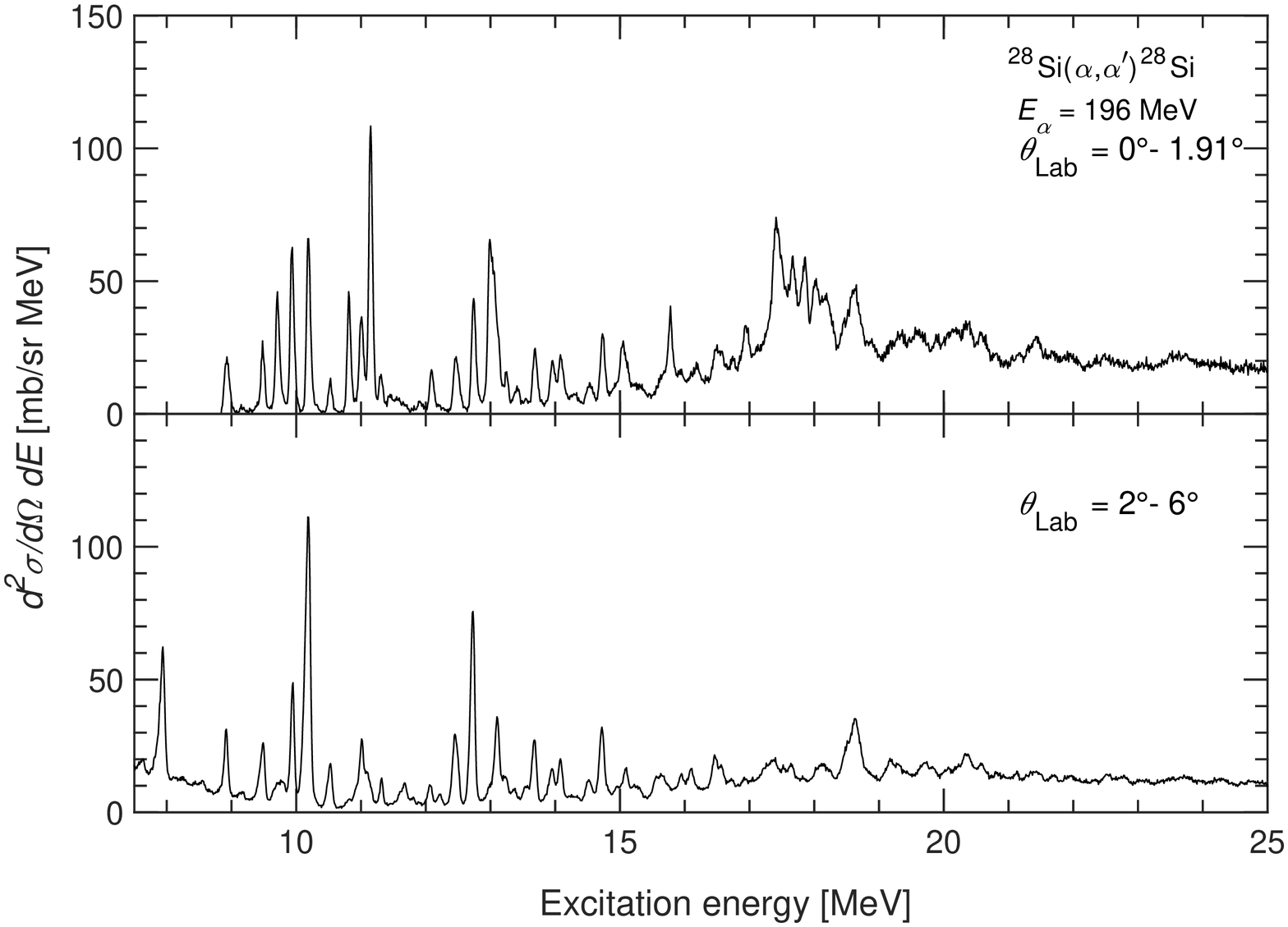}
		\caption{Same as Fig.~\ref{FIG:1a}, but for $^{28}$Si.}
		\label{FIG:1b}
	\end{figure*}
	
\section{Experimental details}
\label{s2}
The experimental procedure followed in this study is similar to that described elsewhere \cite{adsley2021isoscalar,adsley2017alpha}. As such, only salient details are provided here. The experiment was performed at the Separated Sector Cyclotron (SSC) facility of iThemba LABS using a beam of $196$ MeV $\alpha$ particles. Inelastically scattered $\alpha$ particles were momentum analysed by the K$600$ magnetic spectrometer after interacting with either a $0.23$ mg/cm$^2$ thick $^{24}$Mg or a $0.23$ mg/cm$^2$ thick $^{28}$Si foil. The horizontal and vertical positions of the scattered $\alpha$ particles in the focal plane of the spectrometer were measured using two multiwire drift chambers. Energy deposition in plastic scintillators in the focal plane as well as time-of-flight measurements relative to the cyclotron radio frequency were used for particle identification.
	
Spectra were acquired with the spectrometer set at scattering angles $0^{\circ}$ and $4^{\circ}$.
In the former, scattering angles  $\theta_{\text{Lab}} = 0^\circ \pm 1.91^{\circ}$ and in the latter, scattering angles from $\theta_{\text{Lab}} = 2^{\circ}$ to $6^{\circ}$ were covered by a circular spectrometer aperture, respectively. The trajectory of the scattered particles through the focal plane was used to construct the scattering angle.  A multihole collimator was used to calibrate the measured focal-plane angles to scattering trajectories into the spectrometer. 

In the zero-degree mode, both the unscattered beam and the inelastically scattered particles are transported through the K$600$ magnetic spectrometer with the beam exiting the spectrometer only a few centimeters away from the drift chamber position in the high-dispersion focal plane mode of the K$600$. For such measurements, suppression of beam-induced background is critical. In order to subtract the instrumental background, a standard method used at the iThemba LABS K$600$ \cite{neveling2011high} and the RCNP magnetic spectrometers \cite{tamii2009measurement} that exploits differences in the vertical distribution of real and background events was employed. This method establishes background spectra from the regions of the focal plane above and below the vertically focussed band of true events and allows a direct subtraction of these components from the central region of interest. In contrast to the zero-degree mode, improved instrumental background conditions in the small-angle mode allows one to operate the spectrometer in vertical off-focus mode and also use the medium-dispersion focal plane. In the off-focus mode, the vertical position on the focal plane relates to the vertical component of the scattering angle into the spectrometer aperture and allows for its reconstruction \cite{tamii2009measurement}.

The same techniques as employed in Ref.~\cite{adsley2017alpha} were adopted for the analysis of the data. This includes software corrections of kinematic effects and optical aberrations in the horizontal focal-plane position that depend on the vertical focal-plane position and the scattering angle into the spectrometer. The energy calibration was based on well-known states in $^{24}$Mg \cite{kaw2013,bor1981}. An energy resolution of $\approx 70$ keV (FWHM) was attained. Figures \ref{FIG:1a} and \ref{FIG:1b} show the inelastic scattering cross sections extracted at $0^{\circ}-1.91^{\circ}$ and $2^{\circ}-6^{\circ}$ for $^{24}$Mg and $^{28}$Si, respectively. The broad structure seen in Fig.~\ref{FIG:1b} below $10$ MeV is due to hydrogen contaminants in the targets. However, it does not affect the excitation-energy ranges relevant for the extraction of the IS0 strength as demonstrated in the next section.

\section{Experimental Results}
\label{s3}
The Multipole Decomposition Analysis (MDA) technique was employed in numerous studies to extract the IS0 strength distribution in nuclei
\cite{Zabora2021,GC_review2018,gupta2018isoscalar,24Mg_GG_PRC2016,28Si_PGG_PRC2016,24Mg_GG_PLB2015,
patel2014excitation,youngblood2009isoscalar,youngblood2007isoscalar}. However, due to the limited number of angular data points in the present study, the IS0 strength distributions were rather determined via the Difference-of-Spectra (DoS) method \cite{DoSpaper}.
In the DoS method, an excitation-energy spectrum for the angular region associated with the first minimum of the $L = 0$ angular distribution is subtracted from the excitation-energy spectrum taken at $0^{\circ}$. This method, therefore, requires the determination of the suitable angle cut from the measurement at $\theta_{\text{Lab}} = 2^{\circ} - 6^{\circ}$, which is obtained from DWBA calculations.
	
In the present study, the DWBA calculations were performed according to the method described in Ref.~\cite{satchler1997missing}. A density-dependent single-folding model for the real part of the potential $U(r)$, obtained with a Gaussian $\alpha$-nucleon potential, and a phenomenological Woods-Saxon potential for the imaginary term of $U(r)$ were used, so that the $\alpha$-nucleus potential can be written as
\begin{equation}
	\label{e4a}
	U(r) = V_{\text{fold}}(r) + i\dfrac{W}{\left\lbrace 1 + \exp\left[\left(r - R_\text{I}\right)/a_\text{I}\right] \right\rbrace}~,
\end{equation}
with radius $R_\text{I} = r_\text{0I}(A_{\text{p}}^{1/3}+A_{\text{t}}^{1/3})$ and diffuseness $a_\text{I}$. The subscripts p and t refer to projectile and target, respectively, and $A$ denotes the mass number. The potential $V_{\text{fold}}(r)$ is obtained by folding the ground-state density with a density-dependent $\alpha$-nucleon interaction
	\begin{equation}
	\label{e421b}
	V_{\text{fold}}(r) = - V \int d^3r^\prime \rho(r^\prime)\left[1 - \beta\rho(r^\prime)^{2/3}\right]\exp(- \mathit{z}^2/t^2)~,
	\end{equation}
where $\mathit{z} = |r - r^\prime|$ is the distance between the centre of mass of the $\alpha$ particle and a target nucleon, and $\rho(r^\prime)$ is the ground-state density of the target nucleus at the position $r^\prime$ of the target nucleon. The parameters $\beta = 1.9$ fm$^2$ and range $t = 1.88$ fm were taken from Ref.~\cite{satchler1997missing}.
The ground-state density $\rho(r)$ of the target nucleus at the position $r$ is given by\\
	\begin{equation}
	\label{e421c}
	\rho(r) = \dfrac{\rho_0}{1 + \exp\left(\frac{r - c}{a}\right)}~,
	\end{equation}\\
	where the Fermi-distribution parameters $c$ and $a$ describe the half-density radius and the diffuseness, respectively. Numerical values for $^{24}$Mg and $^{28}$Si were taken from Ref.~\cite{fricke1995nuclear}.
	
DWBA calculations were carried out using the computer code PTOLEMY \cite{rhoades1980techniques,rhoades1980techniques2}. The Optical Model Parameters (OMP) used in the DWBA calculations, viz. $r_\text{0I}, a_\text{I}, V$ and $W$, for each nucleus are usually obtained by fitting elastic scattering cross sections. The procedure is fully described in Ref.~\cite{nolte1987global}. However, due to the lack of elastic scattering data in this study, the parameters were taken from studies of the TAMU group on the same nuclei. This is justified by the similarity of their beam energy ($240$ MeV) compared to this study. The OMP parameters are summarised in Table \ref{table:1}.

	\begin{table}
		\caption{Optical model parameters taken from the TAMU group publications.}	
		\label{table:1}
		\begin{center}
			\setlength{\arrayrulewidth}{0.5pt}
			\setlength{\tabcolsep}{0.15cm}
			\renewcommand{\arraystretch}{1.3}	
			\begin{tabular}{cccccc}
				\hline\hline
				Nucleus &  $V$ (MeV) & $W$ (MeV) & $r_\text{0I}$ (fm) & $a_\text{I}$ (fm) & Refs.\\
				\hline
				$^{24}$Mg &  $41.02$  & $35.39$ & $0.934$ & $0.614
				
				$ & \cite{youngblood1999giant}\\
				$^{28}$Si &  $44$  & $32.5$ & $0.9306$ & $0.687
				
				$ & \cite{youngblood2002isoscalar}\\
				\hline
			\end{tabular}
		\end{center}	
	\end{table}

\begin{figure}
		\includegraphics[width=.5\textwidth]
		{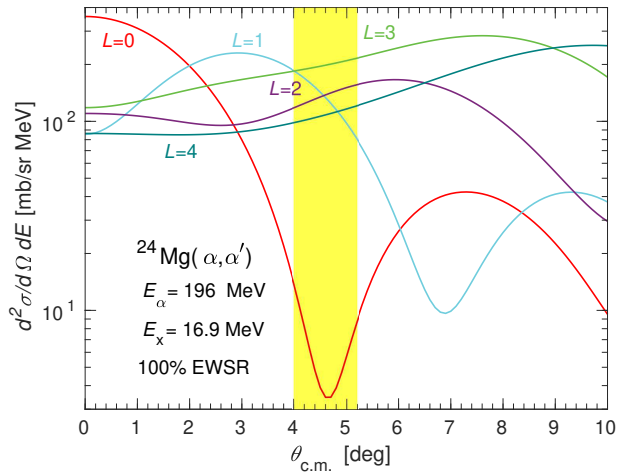}\hfill
		\includegraphics[width=.5\textwidth]
		{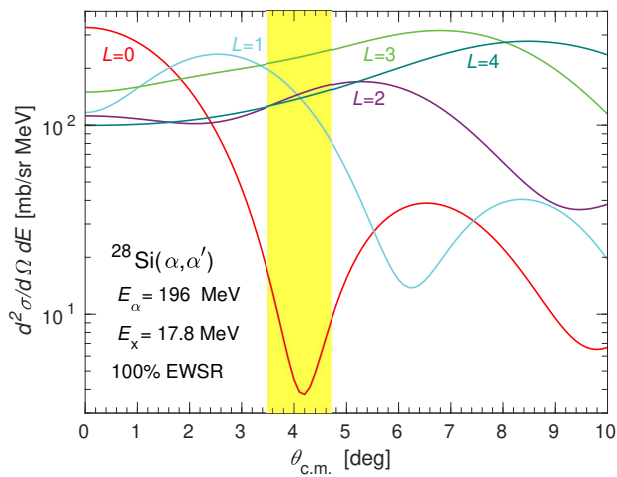}
		
		\caption{DWBA calculations of the differential cross sections for the $^{24}$Mg($\alpha$, $\alpha^\prime$) (top panel) and $^{28}$Si($\alpha$, $\alpha^\prime$) (bottom panel) reaction at $E_\alpha = 196$ MeV for various isoscalar electric multipoles. The calculations have been normalised to $100\%$ of the appropriate EWSR at an excitation energy of $16.9$ MeV in $^{24}$Mg and $17.8$ MeV in $^{28}$Si.}
		\label{FIG: 3}
	\end{figure}

The measured cross sections can be converted to fractions of the EWSR $(a_0)$ by comparing with DWBA calculations assuming $100\%$ EWSR, which are shown in Fig.~\ref{FIG: 3}. The strength is then calculated using the $a_0$ values and is expressed as
	\begin{equation}
	\label{e5}
	S_{0}(E_\text{x}) = \dfrac{2\hbar^2 A \langle r^2\rangle}{mE_\text{x}}a_0(E_\text{x})~,
	\end{equation}
where $m$ is  the nucleon mass, $E_\text{x}$ is the excitation energy corresponding to a given state or energy bin, and $\langle r^2\rangle$ is the second moment of the ground-state density. We use $\langle r^2\rangle = 9.945$ fm$^2$ for $^{24}$Mg and $9.753$ fm$^2$ for $^{28}$Si.
	
The DoS technique relies on the subtraction of the contributions of all multipoles except $L = 0$ from the $0^{\circ}$ spectrum using data from an angle range defined by the first minimum of the $L = 0$ contribution. It was shown that the sum of angular distributions of all multipoles $L > 0$ is similar at the maximum and first minimum of the $L = 0$ distribution \cite{GC_review2018,DoSpaper}.
Hence, the  subtraction of the inelastic spectrum associated with the flat angular distribution (where $L = 0$ is  at a minimum from the $0^{\circ}$ spectrum) is assumed to represent essentially the IS0 component excited in $\alpha$-inelastic scattering close to $0^{\circ}$.
	
\begin{figure}
		\includegraphics[width=.5\textwidth]
		{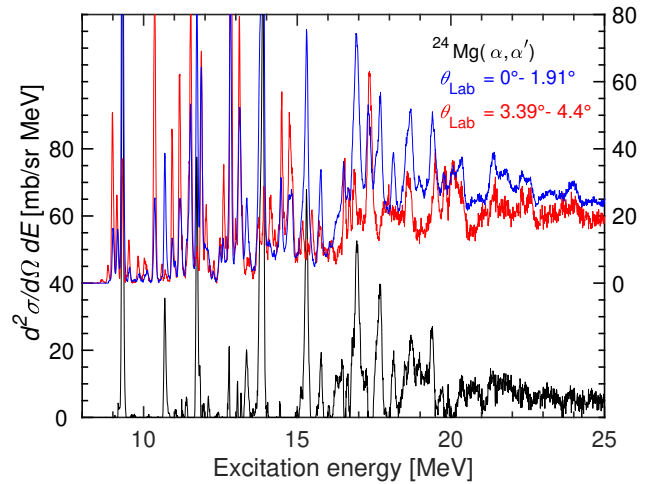}\hfill
		\includegraphics[width=.5\textwidth]
		{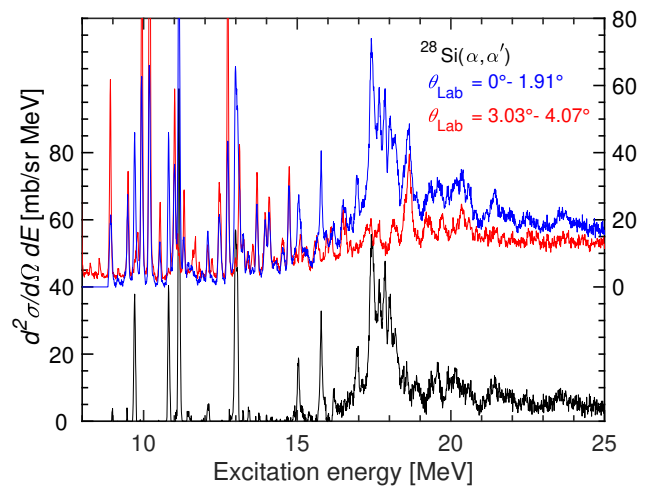}		
		\caption{Double-differential cross sections for $^{24}$Mg($\alpha$, $\alpha^\prime$) (top panel) and $^{28}$Si($\alpha$, $\alpha^\prime$) (bottom panel) at $E_\alpha = 196$ MeV. The blue and red spectra represent the data acquired at $0^{\circ} \leq \theta_{\text{Lab}} \leq 1.91^{\circ}$ and at $3.39^{\circ} \leq \theta_{\text{Lab}} \leq 4.4^{\circ}$ for $^{24}$Mg and $3.03^{\circ} \leq \theta_{\text{Lab}} \leq 4.07^{\circ}$ for $^{28}$Si, respectively. For clarity, these spectra are shifted by $40$ mb/sr MeV. The black spectra represent the difference calculated using the DoS technique (see text).}
		\label{FIG: 5}
	\end{figure}

The direct subtraction of the angle cut spectrum from the spectrum at $0^{\circ} \leq \theta_{\text{Lab}} \leq 1.91^{\circ}$ (blue) yields the black difference spectra shown in the upper and lower panels of Fig.~\ref{FIG: 5} for $^{24}$Mg and $^{28}$Si, respectively. The angle cut for each nucleus is indicated by the highlighted yellow area in Fig.~\ref{FIG: 3} and listed in Table \ref{table:3}. The angle corresponding to the minimum of the $L = 0$ distribution is also indicated.

	\begin{table}
		\caption{Angle cuts implemented in the $4^{\circ}$ dataset to define the angular
                 region around the first $L = 0$ minimum.}	
		\label{table:3}
		\begin{center}
			\setlength{\arrayrulewidth}{0.5pt}
			\setlength{\tabcolsep}{0.3cm}
			\renewcommand{\arraystretch}{2}	
			\scalebox{1}{\begin{tabular}{ccc}
				\hline\hline
				Nucleus & $^{24}$Mg & $^{28}$Si \\
				\hline
				Angle cut ($\theta_{\text{Lab}}$) &  $3.39^{\circ}-4.4^{\circ}$  & $3.03^{\circ}-4.07^{\circ}$		
				\\
				Angle cut ($\theta_{\text{c.m.}}$) &  $4^{\circ}-5.2^{\circ}$  & $3.5^{\circ}-4.7^{\circ}$
				\\
				$L = 0$ minimum ($\theta_{\text{c.m.}}$) &  $4.6^{\circ}$  & $4.1^{\circ}$
				\\
				\hline
				\end{tabular}}
	\end{center}	
	\end{table}	
	
\begin{figure} 
		\centering
		\includegraphics[scale=0.45]{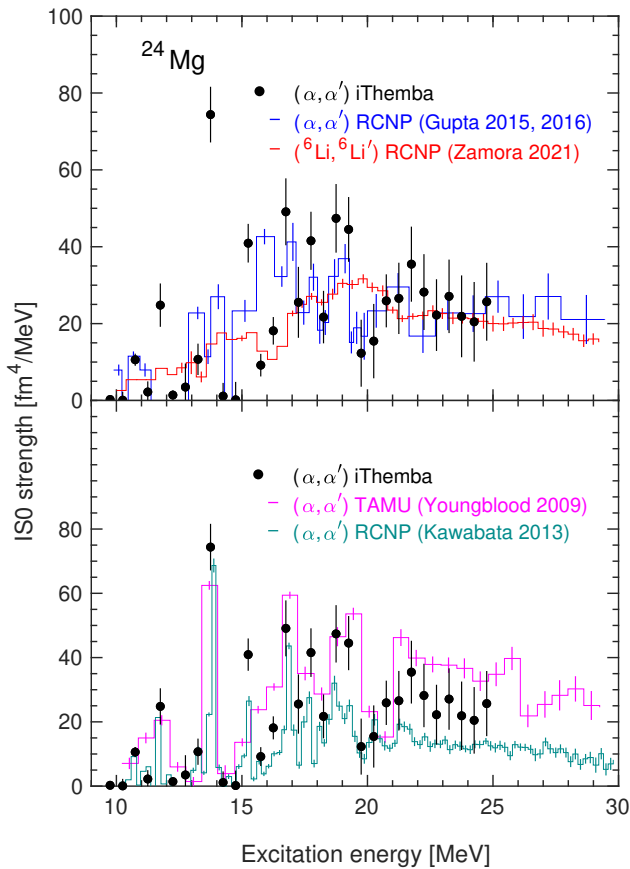}
				\caption{IS0 strength distributions in $^{24}$Mg. The present iThemba LABS data are shown as black filled circles. Also shown are the ($\alpha$, $\alpha^\prime$) \cite{24Mg_GG_PLB2015,24Mg_GG_PRC2016} and ($^{6}$Li, $^{6}$Li$^\prime$) data \cite{Zabora2021} from the RCNP shown as blue and red histograms, respectively (top panel). The bottom panel displays results from TAMU \cite{youngblood2009isoscalar} (magenta histogram) and a different measurement at RCNP \cite{kaw2013} shown by dark cyan histogram.}
		\label{FIG:6}
	\end{figure}

\begin{figure} 
		
		\includegraphics[scale=0.63]{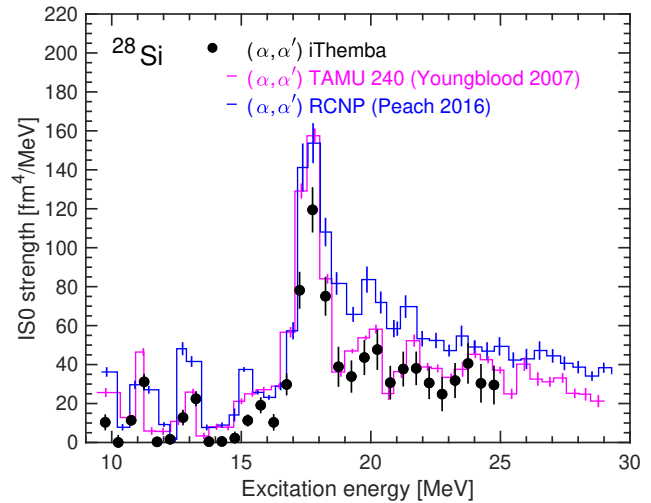}
		
		\caption{IS0 strength distribution in $^{28}$Si. The present iThemba data are shown as black filled circles. Data shown as blue and magenta histograms with vertical error bars are from RCNP \cite{28Si_PGG_PRC2016} and TAMU \cite{youngblood2007isoscalar}.}
		\label{FIG:7}
	\end{figure}

IS0 strength distributions in $^{24}$Mg and $^{28}$Si were determined from the difference spectra. The fraction of IS0 strength in EWSR
per MeV was obtained by dividing the extracted experimental IS0 component by the corresponding integral value of $L = 0$, $100\%$ EWSR obtained at the average angles $\theta_{\text{c.m.}}^{\text{av.}} = 1.13^{\circ}$ ($^{24}$Mg) and $1.12^{\circ}$ ($^{28}$Si) in the $0^{\circ}$ spectra. The results are shown in Figs.~\ref{FIG:6} and \ref{FIG:7} for $0.5$ MeV wide energy bins. The uncertainties shown include both systematical and statistical errors. For comparison purposes, the IS0 strengths from the RCNP experiments  (\cite{kaw2013,24Mg_GG_PLB2015,24Mg_GG_PRC2016} for $^{24}$Mg
and \cite{28Si_PGG_PRC2016}  for $^{28}$Si)
and the TAMU experiments (\cite{youngblood2009isoscalar} for $^{24}$Mg and \cite{youngblood2007isoscalar}  for $^{28}$Si) are also shown. The results from Ref.~\cite{youngblood2009isoscalar} are for the dataset where the out-of-plane angle was measured. For better visibility the comparison of IS0 strength distributions with the present results in Fig.~\ref{FIG:6} for $^{24}$Mg is split into two parts. The upper part shows results from Refs.~\cite{24Mg_GG_PLB2015,24Mg_GG_PRC2016}, \cite{Zabora2021} and the lower those from Refs.~\cite{kaw2013} and \cite{youngblood2009isoscalar}.

Data for $^{24}$Mg from the present study generally agree within error bars with the RCNP and TAMU data. However, above $17$ MeV the IS0 strength from the TAMU group is consistently higher than both the iThemba LABS and RCNP strengths. These discrepancies might be attributed to the background-subtraction method employed by the TAMU group. At the same time, Fig.~\ref{FIG:6} shows that the RCNP \cite{24Mg_GG_PLB2015,24Mg_GG_PRC2016}, TAMU \cite{youngblood2009isoscalar} and present iThemba LABS data for $^{24}$Mg give somewhat different strength distributions below the main ISGMR region. In particular, the RCNP experiments \cite{24Mg_GG_PLB2015,24Mg_GG_PRC2016} suggest the highest IS0 structures at $16-20$ MeV and a two times lower peak at $13-16$ MeV.
A similar structure was observed in the recent ($^{6}$Li, $^{6}$Li$^\prime$) experiment \cite{Zabora2021}. Instead, the TAMU and iThemba LABS data (bottom panel of Fig.~\ref{FIG:6}) give the highest IS0 peak  at about $13.8$ MeV and a broader peak with a maximum at about $18$ MeV. It is remarkable that, unlike the RCNP data \cite{24Mg_GG_PLB2015,24Mg_GG_PRC2016}, another ($\alpha$, $\alpha^\prime$) experiment at RCNP with superior energy resolution \cite{kaw2013} does show the strong narrow peak at $13.8$ MeV (bottom panel of Fig.~\ref{FIG:6}). Moreover, this peak was clearly observed in the early ($\alpha$, $\alpha^\prime$) \cite{Lu86} and ($^{6}$Li, $^{6}$Li$^\prime$) \cite{Denn95} experiments (see the discussion in Sec. $4.2.3$ of Ref.~\cite{Harakeh}). So most of the experimental datasets confirm the existence of the strong narrow IS0 peak at $13.8-13.9$ MeV in $^{24}$Mg. As can be seen in the next section, this peak is of a crucial importance for our present study.
The differences between the various datasets, especially for strong narrow peaks of IS0 strength, can most likely be ascribed to differences in the experimental energy resolution and a subtle interplay between binning effects and differences in
the calibration of the excitation energy.

As for $^{28}$Si (Fig.~\ref{FIG:7}), the iThemba LABS results are mainly consistent with the previous findings by the TAMU group, except for a slightly  lower value of the IS0 strength at the peak of the distribution. The results from RCNP, on the other hand, exhibit almost a factor of two larger IS0 strength above $19$ MeV than both the iThemba LABS and the TAMU results. One should also note that some of the IS0 strength at lower excitation energy might not belong to the ISGMR. It was suggested that, in $^{28}$Si, the $0^+$ states at $9.71$ MeV, $10.81$ MeV, $11.14$ MeV, and $12.99$ MeV, i.e. most of the IS0 strength below $15$ MeV, can be considered as potential band heads for superdeformed bands \cite{adsley2017alpha}.

\begin{table*}[t]
	\caption{Excitation of prominent $J^\pi = 0^+$ states in $^{24}$Mg. $S$(IS0) is the experimental IS0 strength exhausted by the state.}
	\label{table:5}
	\begin{center}
		\setlength{\arrayrulewidth}{0.5pt}
		\setlength{\tabcolsep}{0.4cm}
		\renewcommand{\arraystretch}{2}	
		\begin{tabular}{cccccc}
			\hline\hline
			$E_\text{x}$ (MeV)\footnote{Present experiment.\label{footnote1}}& $E_\text{x}$ (MeV)\footnote{From Ref.~\cite{nndc}.}&  $E_\text{x}$ (MeV)\footnote{From Ref.~\cite{adsley2021isoscalar}.\label{footnote2}}& \% EWSR\footref{footnote1} & \% EWSR\footref{footnote2} & $S$(IS0) (fm$^4$)\footref{footnote1} \\
			\hline
			$9.31(1)$ & $9.30539(24)$ & & $1.0(1)$ & $1.4(3)$  &  $19.9(20)$ 
			\\
			$10.68(1)$& $10.6797(4)$ & & $0.31(5)$ & $0.29(6)$  & $5.32(61)$
			\\
			$11.73(1)$ & $11.7281(10)$ &  & $0.77(11)$ & $1.0(2)$ &$12.1(16)$
			\\
			$13.36(2)$ &  & $13.37(1)$ & $0.40(5)$  & $0.5(1)$ & $4.32(71)$
			\\
			 &  & $13.79(1)$ & & $1.7(3)$ & 
			\\
			
			$13.87(2)\footnote{The integration region for the $13.87(2)$ state from the present experiment includes both the $13.79$ MeV and $13.89$ MeV states from Ref.~\cite{adsley2021isoscalar}.\label{footnote5}}$ & $13.884(1)$ & $13.89(1)$ & $2.8(3)\footref{footnote5}$  & $2.6(5)$ & $37.7(38)\footref{footnote5}$
			\\
			$15.32(2)$ &  & $15.33(3)$ & $1.7(2)$  & $1.9(4)$ & $20.7(25)$ 
			\\
			$15.78(2)$ &  & $15.79(3)$  &$0.40(5)$  & $1.1(2)$ & $4.6(10)$
			\\
			\hline
		\end{tabular}
	\end{center}	
\end{table*}

\begin{table}
	\caption{Same as in Table \ref{table:5} but for excitation of $J^\pi = 0^+$ states in $^{28}$Si.}
	\label{table:6}
	\begin{center}
		\setlength{\arrayrulewidth}{0.5pt}
		\setlength{\tabcolsep}{0.1cm}
		\renewcommand{\arraystretch}{3}	
		\scalebox{0.93}{$\begin{tabular}{ccccc}
			\hline\hline
			$E_\text{x}$ (MeV)\footnote{Present experiment.\label{footnote3}} & $E_\text{x}$ (MeV)\footnote{From Ref.~\cite{adsley2017alpha}.\label{footnote4}} & \% EWSR\footref{footnote3} & \% EWSR\footref{footnote4} & $S$(IS0) (fm$^4$)\footref{footnote3} \\
			\hline
			$9.70(2)$ & $9.71(2)$ &  $0.22(4)$  & $0.38(8)$& $5.2(6)$
			\\
			$10.81(2)$ & $10.81(3)$ &  $0.27(4)$  &$0.35(7)$& $5.7(6)$
			\\
			$11.14(2)$ & $11.142(1)$ &  $0.8(1)$  &$0.9(2)$& $15.3(17)$
			\\
			$13.00(2)$ & $12.99(2)$ &  $0.95(12)$  &$0.8(2)$& $16.9(18)$
			\\
			$15.03(3)$ & $15.02(3)$ & $0.40(9)$  &$0.8(2)$& $6.0(15)$
			\\
			$15.77(3)$ &  & $0.7(1)$ & & $9.7(16)$ 
			\\
			\hline
			\end{tabular}$}
	\end{center}	
\end{table}

With the good energy resolution of the present experiment, $J^\pi = 0^+$ states could be resolved up to $16$ MeV in $^{24}$Mg and 15 MeV in $^{28}$Si. The corresponding strengths $S$(IS0) and \% EWSR exhausted by the strongest discrete states evident in  Fig.~\ref{FIG: 5} are summarised in Tables \ref{table:5} and \ref{table:6}, respectively. The values obtained in this study are in good agreement with results from Ref.~\cite{adsley2021isoscalar} which analysed angular distributions of individual states, employing a different DWBA code as well as OMPs.

	\section{QRPA calculations and comparison with experiment}
	\label{s4}
	
	\subsection{Details of the calculations}
	\label{s41}

The calculations were performed within the QRPA  model
\cite{Rep_arxiv,Rep_EPJA17,Rep_sePRC19,Kva_seEPJA19} based on the Skyrme functional
\cite{Ben_RMP03,Stone_PPNP07}. The model is fully self-consistent. Both the mean field and the residual
interaction are derived from the initial Skyrme functional. The residual interaction takes into
account all the terms following from the Skyrme functional and Coulomb (direct and exchange) parts. Both particle-hole and particle-particle  channels are included \cite{Rep_EPJA17}. Pairing is treated at the BCS level \cite{Rep_EPJA17}. Spurious admixtures caused by pairing-induced violation of the particle
 number are removed using the method from Ref.~\cite{Kva_seEPJA19}.	

A representative set of Skyrme forces is used, see Table \ref{tab-5}. For our aims, the most important characteristics of the forces are the nuclear incompressibility $K_\infty$ and the isoscalar effective mass $m^*_0/m$ affecting ISGMR and ISGQR energies, respectively. The corresponding values are listed in Table \ref{tab-5}. We employ the standard force SkM* \cite{SkM*} and the most recent force SV-bas \cite{SVbas}. Both forces were previously used in the  analysis presented in
\cite{24Mg_GG_PLB2015,24Mg_GG_PRC2016,28Si_PGG_PRC2016,24Mg_PPJC2016}. Further, we use the earlier force  SkT6 \cite{SkT6} and more recent force
SV-mas10 \cite{SVbas} which both have a large effective mass, $m^*_0/m = 1$, and so are
favourable for the description of the ISGQR \cite{E0_Colo_PLB2020,Ne_odd,Kureba}.
Finally, we exploit the force SkP$^{\delta}$ \cite{SkP} which has a very low incompressibility
$K_\infty = 202$ MeV. In a recent study of the ISGMR in Mo isotopes, this force gave the best results \cite{E0_Colo_PLB2020}.

\begin{figure} 
		\centering
		\includegraphics[width=8cm]{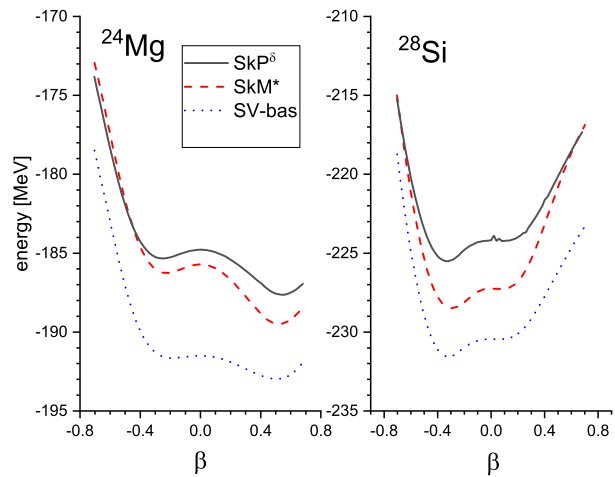}
        \vspace{0.5cm}
\caption{Total energy of $^{24}$Mg  (left) and $^{28}$Si (right), calculated
in the framework of  QRPA with the forces SkP$^{\delta}$, SkM* and SVbas as a
function of the deformation parameter $\beta$.}
\label{FIG:7ensurf}
\end{figure}

The nuclear mean field and pairing are computed by the code SKYAX \cite{Skyax} using a two-dimensional
grid in cylindrical coordinates. The calculation box extends up to three nuclear radii and the grid step size is $0.4$ fm.  The axial quadrupole equilibrium deformation is obtained by minimisation of the energy of the system. As shown in Fig.~\ref{FIG:7ensurf}, both $^{24}$Mg and $^{28}$Si  have oblate and prolate minima. However, in the ground state, $^{24}$Mg is prolate and $^{28}$Si is oblate. As shown in Table \ref{tab-6}, the calculated deformation parameters $\beta$ somewhat underestimate the experimental values \cite{nndc}. This is typical for light deformed nuclei, see e.g. Ref.~\cite{Ne_PRL18}. For all the applied Skyrme forces, the pairing in  $^{24}$Mg and $^{28}$Si was found to be very weak. This is explained by the sparse single-particle spectrum in light nuclei.

	\begin{table} [t]
		
		\caption{Incompressibility $K_\infty$ and isoscalar effective mass $m^*_0/m$
			for the Skyrme forces   SV-bas, SkM*, SkP$^{\delta}$, SkT6 and SV-mas10 used in the present analysis.}
		\begin{center}
		\setlength{\arrayrulewidth}{0.5pt}	
		\label{tab-5}       
		\begin{tabular}{cccccc}
			\hline\hline
			& SV-bas & SkM* & SkP$^{\delta}$ & SkT6 & SV-mas10 \\
			\hline
			$K_\infty$ (MeV)   &  $234$  & $217$ &  $202$ & $236$   & $234$  \\
			$m^*_0/m$  & $0.9$   & $0.79$ & $1$ & $1$ & $1$ \\
			\hline
		\end{tabular}
	\end{center}
	\end{table}
	
The calculations use a large configuration space.
The single-particle spectrum extends from the bottom of the potential well up to $30$ MeV. For example, SkP$^{\delta}$ calculations use $404$ proton and $404$ neutron single-particle levels for $^{24}$Mg and $422$ proton and $422$ neutron single-particle levels for $^{28}$Si. The two-quasiparticle ($2$qp) basis in QRPA calculations with SkP$^{\delta}$ extends up to $70$ MeV.
It includes $1056$ proton and $956$ neutron $K^{\pi} = 0^+$ states in $^{24}$Mg and $1142$ proton
and $1002$ neutron states in  $^{28}$Si. The isoscalar monopole energy-weighted sum rule
EWSR(IS0)$ = 2\hbar^2/m \cdot A \langle r^2 \rangle_0$  is exhausted by $97 - 99$\%.
	
QRPA strength functions for isoscalar monopole $(L = 0)$ and quadrupole $(L = 2)$ transitions have the form
\begin{equation}
	\label{eq:strength}
	S_L(E) = \sum_{K=0}^L (2-\delta_{K,0}) \sum_{\nu \in K}
	\vert \langle \nu \vert {\hat O}_{LK} \vert 0 \rangle \vert^2
	\xi_{\Delta}(E-E_\nu),
\end{equation}
where $\nu$ labels QRPA eigenvalues $|\nu\rangle$ with energies $E_{\nu}$, and $|0\rangle$
is the QRPA ground state. The monopole and quadrupole isoscalar transition operators are $\hat O_{00} = \sum_i^A r^2_i $ and $\hat O_{20} = \sum_i^A r^2_i Y_{20} (\hat r_i)$, respectively. For convenient comparison with the experimental data, the strength is smoothed by a Lorentz function $\xi_{\Delta}(E-E_{\nu}) = \Delta /(2\pi[(E-E_{\nu})^2 - \Delta^2/4])$ with an averaging parameter $\Delta = 1$ MeV. The dimension of the strength functions is fm$^4$/MeV.
We also present 2qp strength functions calculated without the residual interaction. In this case,
the number $\nu$ in Eq. (\ref{eq:strength}) labels 2qp states.

\begin{table} [t]
		\caption{Experimental and calculated deformation parameters $\beta$ in $^{24}$Mg
			and  $^{28}$Si.}
			\begin{center}
			\setlength{\arrayrulewidth}{0.5pt}	
		\label{tab-6}       
		\begin{tabular}{ccccccc}
			\hline\hline
			& Exp.\footnote{From Ref.~\cite{nndc}.} & SV-bas & SkM* & SkP$^{\delta}$ & SkT$6$ & SV-mas$10$ \\
			\hline
			$^{24}$Mg & $0.613$ & $0.527$ & $0.522$ & $0.545$ & $0.506$ & $0.518$ \\
			$^{28}$Si & $-0.412$ & $-0.308$ &  $-0.291$ & $-0.322$ & $-0.311$ & $-0.295$ \\
			\hline
		\end{tabular}
	\end{center}
	\end{table}

\begin{figure*} 
		\centering
		\includegraphics[height=9.5cm,width=12cm]{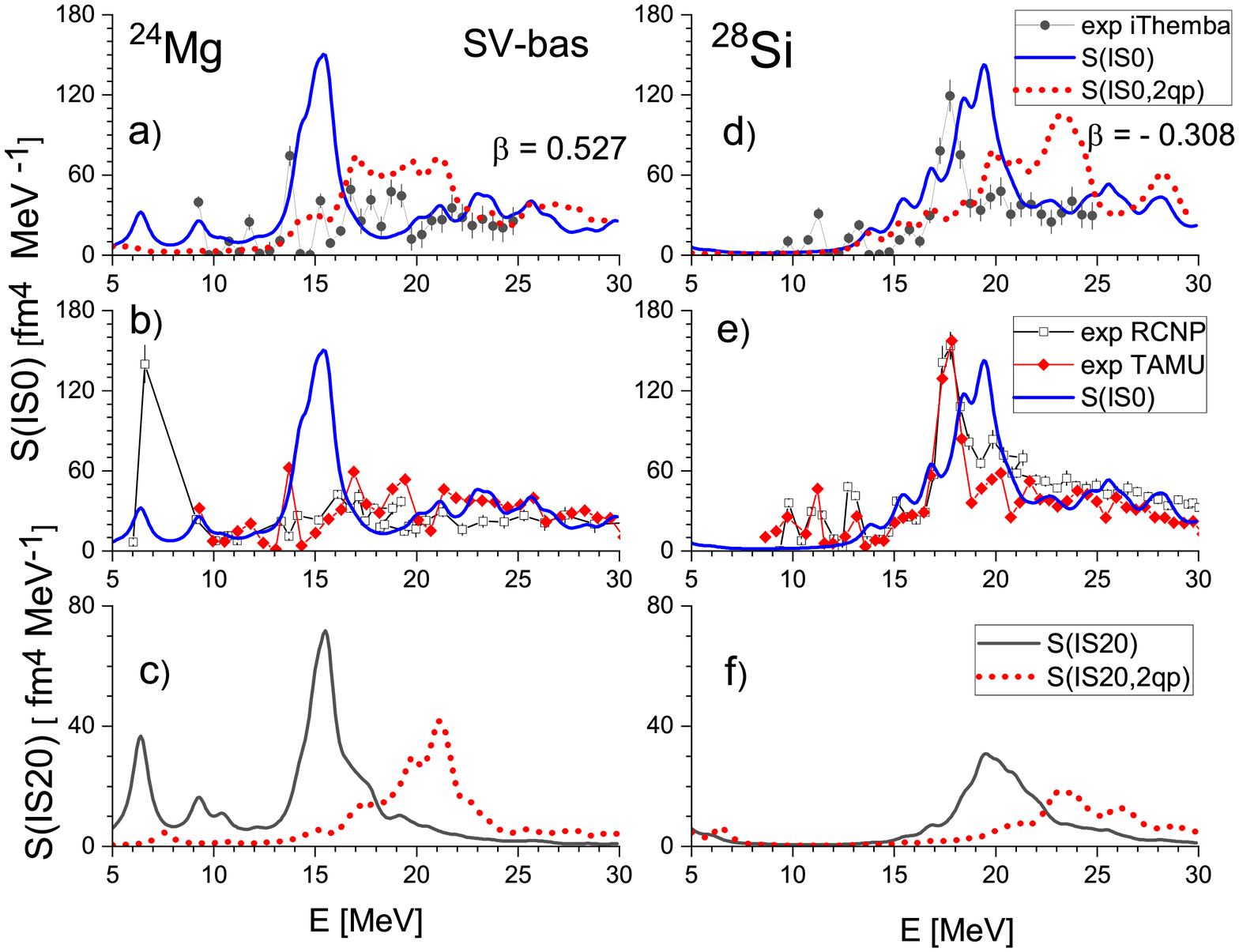}
		\caption{Upper panels: QRPA (solid black line) and 2qp (short-dash red line) IS0
strength functions calculated with the force SVbas ($K_\infty = 234$ MeV) in $^{24}$Mg (left)
and $^{28}$Si (right) and comparison with the present iThemba LABS experimental data shown as
black filled circles. Middle panels: comparison with RCNP \cite{24Mg_GG_PRC2016,28Si_PGG_PRC2016}
(black empty squares) and TAMU \cite{youngblood2009isoscalar,youngblood2007isoscalar}
(red filled diamonds) experimental data. Bottom panels: QRPA (solid black line) and 2qp
(short-dash red line) IS20 strengths.}
		\label{FIG:8_SVbas}
	\end{figure*}
	
\subsection{Strength distribution and IS0/IS2 coupling}
\label{s42}
	
\begin{figure} 
		\centering
		\includegraphics[height=7.5cm,width=9cm]{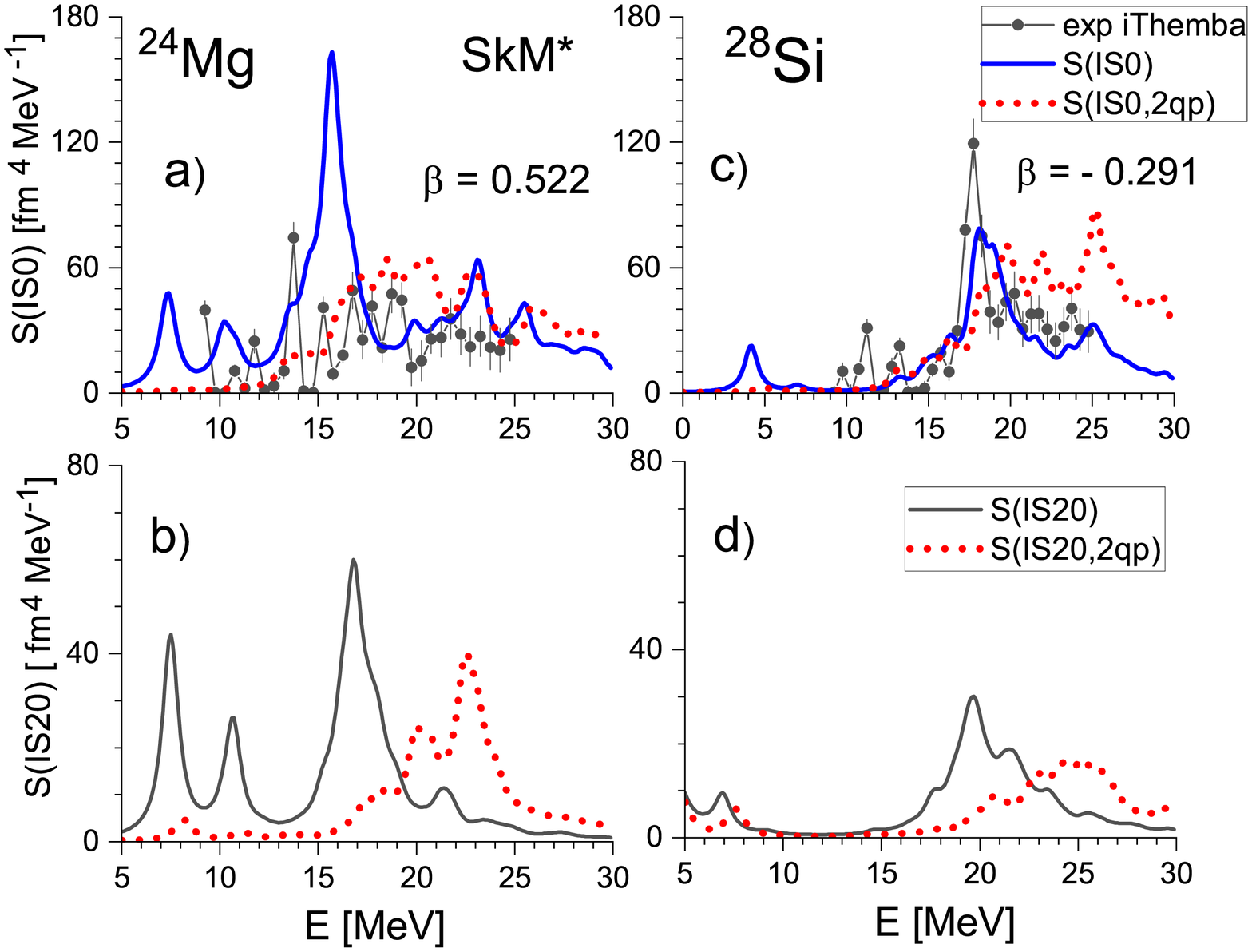}
		\caption{Same as upper and lower panels of Fig.~\ref{FIG:8_SVbas} but for the force SkM* with $K_\infty = 217$ MeV.}
		\label{FIG:8_SkMs}
	\end{figure}

\begin{figure} 
		\centering
		\includegraphics[height=7.5cm,width=9cm]{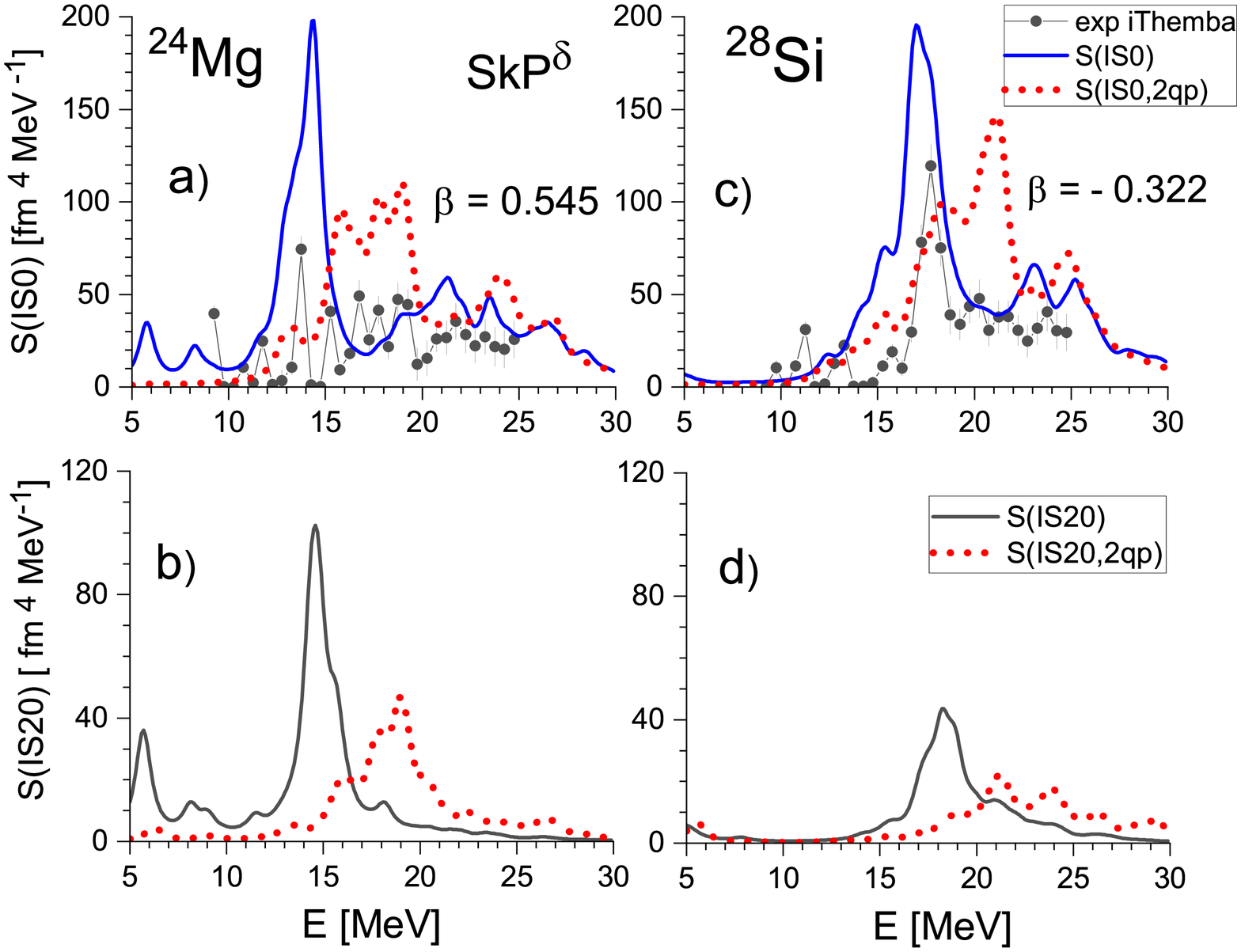}
		\caption{Same as in Fig.~\ref{FIG:8_SkMs} but for the force
SkP$^{\delta}$  with $K_\infty = 202$ MeV.}
		\label{FIG:9_SkP}
	\end{figure}

\begin{figure}[t] 
		\centering
		\includegraphics[height=7cm,width=8cm]{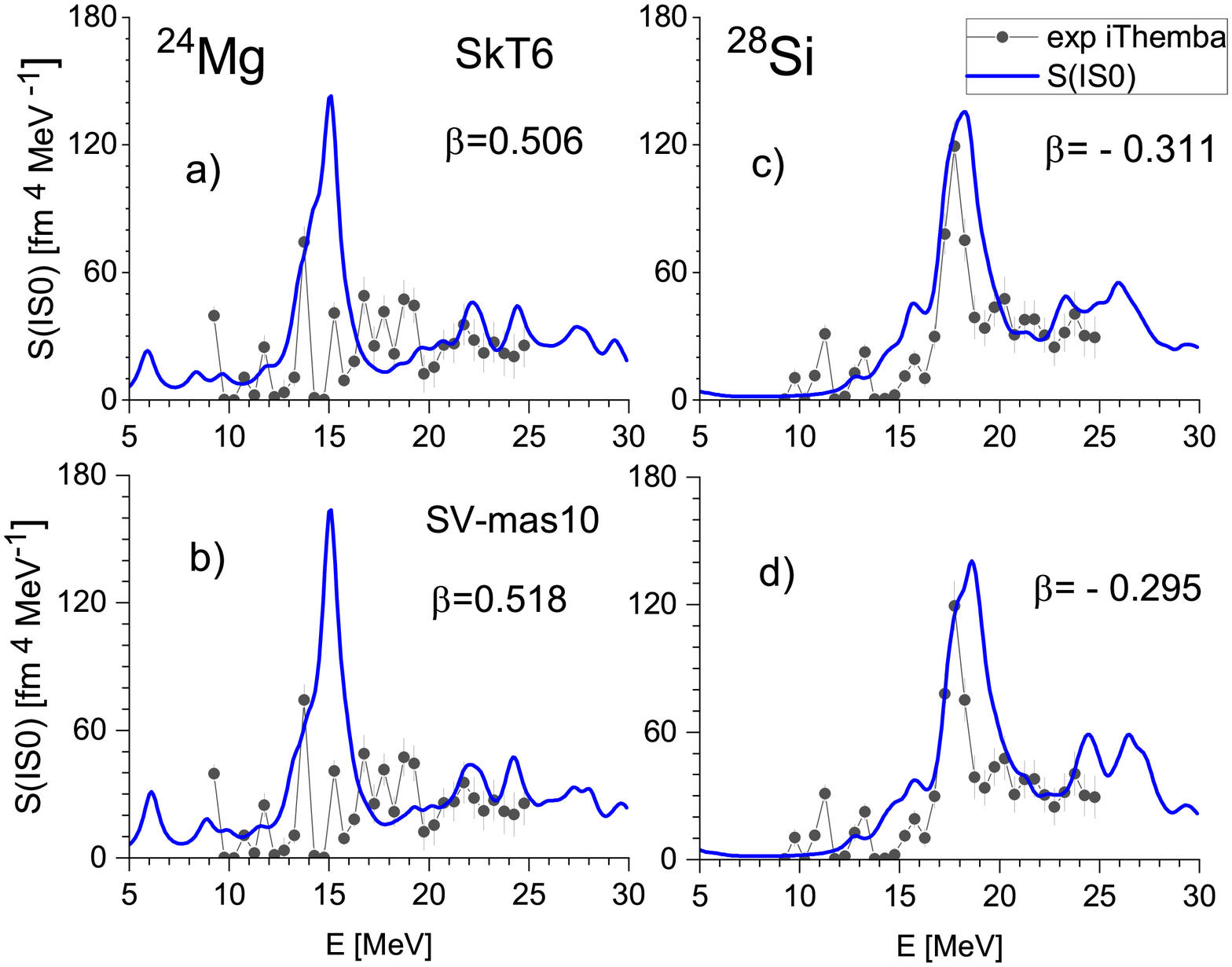}
		\caption{QRPA IS0 strength functions (solid lines) calculated with the Skyrme forces SkT6 (upper panels) and SV-mas10 (bottom panels).}
		\label{FIG:11_SkT6}
	\end{figure}

The QRPA IS0 strength functions from Eq. (\ref{eq:strength}) are compared with the
experimental data in Figs.~\ref{FIG:8_SVbas} - \ref{FIG:11_SkT6}. The iThemba LABS
experimental data are shown by black filled circles where each data point accumulates
the IS0 strength in an energy interval of $0.5$ MeV.
	
In Fig.~\ref{FIG:8_SVbas}, the strength functions calculated with the force SV-bas are shown.
In our set of Skyrme forces, SV-bas has a large (though rather typical for Skyrme forces)
incompressibility $K_\infty = 234$ MeV. The
calculated QRPA IS0 strengths (blue solid line) in $^{24}$Mg and $^{28}$Si are separated
into a narrow structure at $\sim 14$ MeV in $^{24}$Mg and $\sim 18$ MeV in $^{28}$Si and
a broad structure at higher  energy, which is a typical picture for deformed nuclei
\cite{E0_Kvasil_PRC2016}. It is instructive to compare QRPA IS0 strength with the 2qp IS0 one (red dashed line) obtained without the residual interaction. We see a strong collective effect: the residual interaction essentially shifts down the IS0 strength and creates collective peaks at $\sim 14$ MeV in $^{24}$Mg and $\sim 18$ MeV in $^{28}$Si.

The bottom panels of Fig.~\ref{FIG:8_SVbas} show the $LK = 20$ branches of the ISGQR (see Appendix \ref{s6}
for discussion of the deformation splitting of the ISGQR in prolate and oblate nuclei) in the QRPA and 2qp cases. Here, we also observe a strong collective effect. Additionally, one observes that the main IS20 QRPA peak lies at the same energy as the narrow IS0 QRPA peak. Thus, both peaks are related, i.e., we observe a clear deformation-induced coupling of the IS0 and IS20 modes. The larger the nuclear deformation, the more strength is concentrated
in the narrow IS0 peak \cite{E0_Kvasil_PRC2016}. Altogether, we see that the narrow IS0 peaks
in  $^{24}$Mg and $^{28}$Si have a dual  (collective effect + deformation-induced IS0/IS20 coupling) origin.

Now the question arises which structures in the experimental data should be associated
with the calculated IS0 narrow peak. To inspect this point, let us consider in Fig.
\ref{FIG:8_SVbas} the experimental data from iThemba LABS (upper panel), RCNP  \cite{24Mg_GG_PRC2016,28Si_PGG_PRC2016} and TAMU
\cite{youngblood2009isoscalar,youngblood2007isoscalar} (middle panel). In $^{28}$Si, all the data give a distinctive strong narrow peak at $17-19$ MeV. In
$^{24}$Mg, the issue  is more complicated. Here, the iThemba LABS and TAMU data give a distinctive narrow peak at $13-14$ MeV but the RCNP data \cite{24Mg_GG_PRC2016} do not. At the same time, a hump at $13-14$ MeV also appears in the ($^{6}$Li, $^{6}$Li$^\prime$) data \cite{Zabora2021}. Moreover, it is clearly observed at $13.8$ MeV in the RCNP ($\alpha$, $\alpha^\prime$)  data of Ref.~\cite{kaw2013}, which have an energy resolution comparable to the present experiment. Finally,
the narrow peak  at $13-14$ MeV was observed in early ($\alpha$, $\alpha^\prime$)
\cite{Lu86} and ($^{6}$Li, $^{6}$Li$^\prime$) \cite{Denn95} experiments. So, in $^{24}$Mg, the peak at $13-14$ MeV is a good candidate for the comparison with the narrow IS0 peak in our QRPA calculations and the analysis of IS0/IS2 coupling.

Note that the proper choice of the experimental IS0 peak for the comparison  with theory is of crucial importance in the present study. As argued above, this peak arises due to IS0/IS2 coupling in deformed nuclei and its energy location can be used to find the optimal Skyrme force and thereby determine the relevant incompressibility $K_\infty$ and isoscalar
effective mass $m^*_0/m$. In Ref.~\cite{24Mg_GG_PRC2016}, the local maximum at $16-17$ MeV was assumed to be the narrow peak, which resulted in the preference of  the force SkM* ($K_\infty = 217$ MeV, $m^*_0/m = 0.79$). Below we will show that in $^{24}$Mg, the experimental peak at $13-14$ MeV in the iThemba LABS and TAMU data is a better suited candidate. This will change the predicted values of  $K_\infty$ and $m^*_0/m$.

To justify the latter, let us compare the calculated and experimental IS0 strengths in Fig. \ref{FIG:8_SVbas}. It is seen that the calculated narrow IS0 peaks lie higher by $1.5 - 2$ MeV than
the experimental peaks at $13.8$ MeV ($^{24}$Mg) and about $18$ MeV ($^{28}$Si). If we additionally upshift the calculated IS0 strength in both nuclei, we could match the RCNP structure at $16-17$ MeV in $^{24}$Mg but, simultaneously, would essentially worsen the description of the strength in $^{28}$Si. Instead, if we assume  the $13.8$ MeV peak as the relevant narrow structure in $^{24}$Mg, then it remains possible to describe IS0 results in $^{24}$Mg and $^{28}$Si consistently by an equal downshift of the QRPA narrow peaks.

Figure \ref{FIG:8_SVbas} shows that SV-bas generally well describes the main ISGMR
located at $16-25$ MeV in $^{24}$Mg and $20-25$ MeV in $^{28}$Si. The only noticeable discrepancy
is that the theory noticeably underestimates IS0 strength at $16-20$ MeV in $^{24}$Mg where the experiment shows a sequence of prominent peaks. This discrepancy can be caused by the disregard in our calculations of the Coupling with Complex Configurations (CCC). The CCC could redistribute the strength, reducing the height of the narrow IS0 peak
and shifting a part of its strength to the range $16-20$ MeV.

Previous work has shown that the lower $K_\infty$, the smaller the ISGMR centroid energy \cite{GC_review2018,E0_Kvasil_PRC2016}. In Fig.~\ref{FIG:8_SVbas}, SV-bas gives IS0 narrow peaks at too high excitation energy. So it is worth to try a force with a smaller $K_\infty$. In Fig.~\ref{FIG:8_SkMs}, results for the SkM* force with $K_\infty = 217$ MeV are shown. As in Fig.~\ref{FIG:8_SVbas},  we see a
strong collective effect. The energy of the narrow IS0 peak again coincides with the energy of the quadrupole $LK = 20$ branch, which means that this peak is produced by the IS0/IS2 coupling. However, SkM* does not improve the results: the energy of the IS0 narrow peak is still noticeably overestimated. The downshift of the IS0 strength due to the smaller $K_\infty = 217$ MeV is countered by an upshift due to the smaller value of the effective mass $m^*_0/m = 0.79$ (see the discussion of the impact of the effective mass, in connection with Fig.~\ref{FIG:11_IS2_SkMs_SVbas_SkP}).

Next, in Fig.~\ref{FIG:9_SkP}, we present results for the force SkP$^{\delta}$ with
a low incompressibility $K_\infty = 202$ MeV and high effective mass $m^*_0/m = 1$.
Now both factors downshift the IS0 strength as compared with the SV-bas case. An
improved agreement with the experimental IS0 strength distributions is observed: both in
$^{24}$Mg and $^{28}$Si the calculated IS0 narrow peaks now match the experimental structures.
Note that SkP$^{\delta}$ also provides the best description of the ISGMR in deformed Mo isotopes
\cite{E0_Colo_PLB2020}. In Fig.~\ref{FIG:9_SkP}, the narrow peaks are again produced by
IS0/IS2 coupling and a strong collective effect takes place.
	
In Figs.  \ref{FIG:8_SVbas} - \ref{FIG:9_SkP}, the quadrupole IS20 peak in $^{24}$Mg is stronger and has a lower energy than in $^{28}$Si. As a result, the narrow IS0 peak in $^{24}$Mg is more separated from the main ISGMR than in $^{28}$Si. This is explained by the larger absolute quadrupole deformation of $^{24}$Mg as compared with $^{28}$Si, which results in a stronger downshift of the IS20 peak in $^{24}$Mg. Besides, Figs.  \ref{FIG:8_SVbas} - \ref{FIG:9_SkP} show that IS0/IS2 coupling is relevant also for lower-energy states.

Due to the IS0/IS2 coupling, the energy of the narrow peak is sensitive to the isoscalar
effective mass $m^*_0/m$ which affects the ISGQR energy \cite{E0_Colo_PLB2020,Ne_odd}. In general,
the larger  $m^*_0/m$, the smaller the ISGQR centroid energy \cite{Ne_odd}. So, in principle, one can
try to downshift the energy of the narrow IS0 peak solely using a large value of $m^*_0/m$.
In Fig.~\ref{FIG:11_SkT6}, the forces SkT6 ($K_\infty = 236$ MeV) and SV-mas10 ($K_\infty = 234$ MeV),
which both have an effective mass $m^*_0/m = 1$, are applied for the calculation of the IS0 strength. It is seen  that these two forces still overestimate the energies of the narrow IS0 peaks like in the cases of SV-bas and SkM*. Thus, the optimum Skyrme force also needs a lower value of $K_\infty$.

 We note that a QRPA calculation of the ISGMR strength in $^{24}$Mg with a Gogny force is discussed in Ref.~\cite{Zabora2021}. There, a reasonable agreement with the data could only be achieved after an ad-hoc upshift of the IS0 strength distribution by $2$ MeV.

\subsection{IS2 strength}

Since the energy  of the narrow monopole peak is determined by the $K = 0$ branch of the ISGQR,
it is important to consider the ISGMR and ISGQR simultaneously. In particular, we should  see how well we can describe the ISGQR and its deformation splitting in $^{24}$Mg and $^{28}$Si. This is investigated in Appendix \ref{s6} for various values of the deformation parameter $\beta$. For $^{24}$Mg, only prolate deformations are inspected. For $^{28}$Si, where oblate and prolate deformations for excited states were predicted \cite{adsley2017alpha,Taniguchi_PRC2009,Taniguchi_PLB2020,Afan+Ray}, both types of deformations are considered. In $^{24}$Mg, the calculations show the behaviour expected for prolate deformation splitting where the energies of the branches $LK =  20, 21,$ and $22$ follow the relation $E_{20} < E_{21} < E_{22}$. The branch $LK =  20$ is strong. It is downshifted by $4 - 5$ MeV when increasing the deformation from $\beta = 0.1$ to $0.522$. The same splitting features are seen in prolate $^{28}$Si. However, in oblate $^{28}$Si,
the deformation splitting of ISGQR looks surprising. The branch $LK = 20$ is not upshifted,
as would be expected, but downshifted (by $3 - 4$ MeV from $\beta = -0.1$ to $-0.412$) and
becomes rather weak.

 \begin{figure} 
		\centering
		\includegraphics[height=7.5cm,width=8.5cm]{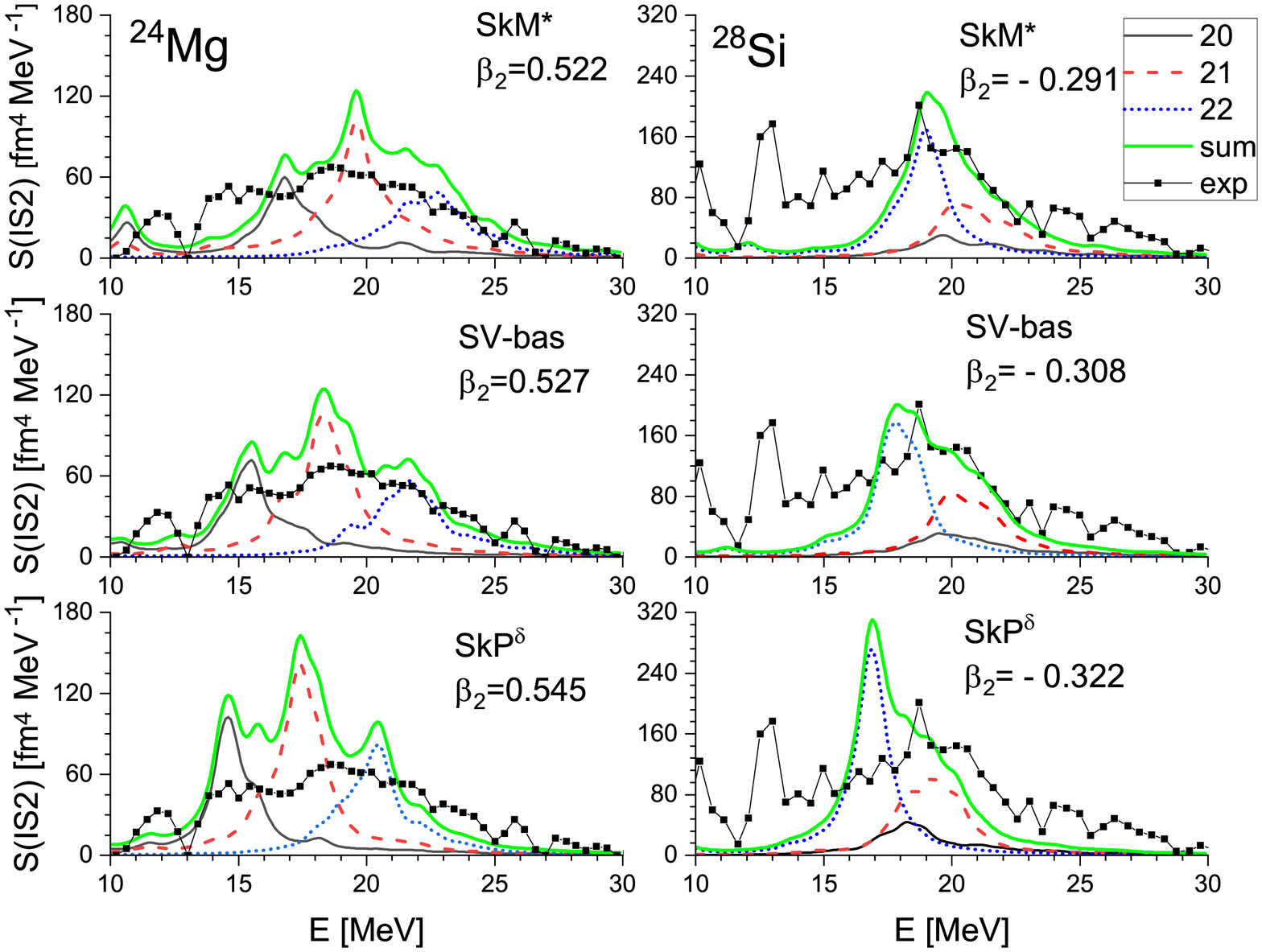}
		\caption{ISGQR in $^{24}$Mg (left) and $^{28}$Si (right), calculated with the forces SkM* (upper panels), SV-bas (middle panels) and SkP$^{\delta}$ (bottom panels). Energy-weighted strength functions $S (\text{IS2})$ for the branches $LK = 20$ (thin solid black line), $21$ (dashed red line) and $22$ (dotted blue line) as well as for  the summed $20$+$21$+$22$ strength (bold green line) are shown. The experimental data are taken from Ref.~\cite{Zabora2021} for $^{24}$Mg and Ref.~\cite{youngblood2007isoscalar} for $^{28}$Si. For better visibility, the latter are multiplied by a factor of 1$0$.}
		\label{FIG:11_IS2_SkMs_SVbas_SkP}
\end{figure}
	
These features of the deformation splitting are illustrated in Fig.~\ref{FIG:11_IS2_SkMs_SVbas_SkP}, where SkM*, SV-bas and SkP$^{\delta}$ QRPA results are compared with ISGQR data obtained in the reactions
($^{6}$Li, $^{6}$Li$^\prime$) for $^{24}$Mg \cite{Zabora2021} and ($\alpha$, $\alpha^\prime$) for $^{28}$Si \cite{youngblood2007isoscalar}. For better visibility, the experimental values for $^{28}$Si are multiplied by a factor of $10$. All three forces reasonably reproduce the IS2 experimental data. In $^{24}$Mg, the best description  is obtained with SV-bas. In $^{28}$Si, SkM* and SV-bas give reasonable results. It is seen that the humps of the experimental strength at $14$ and $19$ MeV in $^{24}$Mg are produced by the $LK = 20$ and $21$ branches, respectively. In $^{28}$Si, the main peak at $19$ MeV is produced by the $LK = 22$ branch. Figure \ref{FIG:11_IS2_SkMs_SVbas_SkP} demonstrates a regular dependence of the ISGQR features on increasing the isoscalar effective mass $m^*_0/m$ from $0.79$ (SkM*) to $0.9$ (SV-bas) to $1.0$ (SkP$^{\delta}$). The larger the value of $m^*_0/m$, the stronger the deformation (and the deformation splitting) and the larger the energy downshift of the ISGQR. The force SkP$^{\delta}$, being best in description of the ISGMR, turns out to be not very good for the ISGQR. A consistent description of the ISGMR and ISGQR would need a Skyrme force with $m^*_0/m\approx 0.9$ and $K_\infty < 217$ MeV (the same conclusion was drawn earlier in the study of ISGMR/ISGQR in Mo isotopes \cite{E0_Colo_PLB2020}). The forces used above do not fulfil these requirements.

Note that our QRPA IS2 strength functions are computed for band heads
$I^{\pi}K = 0^+0, 1^+1, 2^+2$ while IS2 experimental data are obtained for the
states $I^{\pi}K = 2^+0, 2^+1, 2^+2$, i.e. for rotational states in $K = 0$ and $1$ cases.
So, for a more accurate comparison with the experiment, the calculated IS2 $(K = 0,1)$
strengths should be shifted to a higher energy by the rotational correction. It can be estimated roughly from the energy of the $2^+0$ state in the ground-state rotational band, which is $1.368$ MeV for $^{24}$Mg and $1.779$ MeV for $^{28}$Si \cite{nndc}. Though the corrections are significant, they do not affect the conclusions drawn from Fig.~\ref{FIG:11_IS2_SkMs_SVbas_SkP}.
        \begin{table} [b]
		\caption{Peak energy $E_\text{p}$ of the narrow IS0 resonance and summed strength
            $\sum_{\nu} B_{\nu}(\text{IS0})$ in the energy interval $9 - 25$ MeV from
            various Skyrme forces compared with the present experimental data.}
        	\begin{center}

   \vspace{0.2cm}
		\label{tab-7}       
		\setlength{\arrayrulewidth}{0.5pt}
		\setlength{\tabcolsep}{0cm}
		\renewcommand{\arraystretch}{2}	
		\scalebox{0.94}{$\begin{tabular}{cccccc}
			\hline\hline
                    &  & \multicolumn{2}{c}{$^{24}$Mg} & \multicolumn{2}{c}{$^{28}$Si} \\
            \hline
                    &  $K_\infty$ & $E_\text{p}$ \;& $\sum B(\text{IS0})$ & $E_\text{p}$ \;& $\sum B(\text{IS0})$ \\
                    &    \;(MeV) \;   & \;(MeV)\;  & \;(fm$^4$)\;  & \;(MeV)\;  & \;(fm$^4$)\; \\
			\hline
            \; Exp. \; &             & 13.75(2)  & 728(41) & 17.75(3) & 895(40)\\
             \; SkP$^{\delta}$ \;& 202  & 14.3  & 796  & 17.0  & 908\\
            \; SkM* \; &  217          & 15.6  & 706  & 18.0 &  780 \\
            \; SVbas \;&  234          & 15.4 & 634 &  19.4  & 685 \\
            \hline
			\end{tabular}$}
		\end{center}
	\end{table}	

        \begin{table*}

		\caption{Energy-weighted strengths  $\sum_{\nu} E_{\nu}B_{\nu}(\text{IS0})$
                 for $^{24}$Mg and $^{28}$Si,  summed in the energy intervals
                $0-16$ MeV, $16-25$ MeV, $25-70$ MeV, and $0-70$ MeV. The strengths are given in $\%$ of the monopole EWSR (see text for more detail). In the interval $0-16$ MeV,
                 the experimental strength is measured only in the interval $9-16$ MeV.}
   \begin{center}
		\label{tab-8_EWSR}       
		\setlength{\arrayrulewidth}{0.5pt}
		\setlength{\tabcolsep}{0.3cm}
		\renewcommand{\arraystretch}{2}	
		\begin{tabular}{ccccccccc}
			\hline\hline
                    &  \multicolumn{4}{c}{$^{24}$Mg} & \multicolumn{4}{c}{$^{28}$Si} \\
            \hline
                    & \;  $0-16$ & \; $16-25$  & \; $25-70$  & \; $0-70$
                    & \; $0-16$ & \; $16-25$ & \; $25-70$  & \; $0-70$\\
			\hline
             \; Exp.          & 15(1) & 55(4) & 0 & 70(4)&  7(1)  & 68(4) & 0 &75(4)  \\
             \; SkP$^{\delta}$ & 38 & 35 & 25 & 98 & 11 & 63 & 25 & 97 \\
             \; SkM*            & 31 & 37 & 29 & 97 &  5 & 64 & 29 & 98  \\
             \; SV-bas         & 35 & 25 & 37 & 97 &   6 & 55 & 38 & 99\\
            \hline
			\end{tabular}
		   \end{center}
	\end{table*}

\subsection{Integral IS0 strength}

For further comparison, it is worth considering the integrated IS0 strengths
$\sum_{\nu} B_{\nu}(\text{IS0})$ in the energy range $9 - 25$ MeV covered by the present experiment. Here,
$B_{\nu}(\text{IS0}) = \vert \langle \nu \vert {\hat O}_{2K} \vert 0 \rangle \vert^2$ is the reduced probability of the IS0 transition from the ground to $\nu$-th QRPA state. The results for the main three forces used in the present study (SkP$^{\delta}$, SkM*
and SV-bas) are shown in Table \ref{tab-7}. One observes an anticorrelation between $K_\infty$ and $\sum_{\nu} B_{\nu}(\text{IS0})$: the calculated  $\sum_{\nu} B_{\nu}(IS\text{0})$ steadily decreases from  SkP$^{\delta}$ to SV-bas, i.e. with increasing
$K_\infty$ from $202$ MeV to $234$ MeV. As demonstrated for the strength distributions in Figs.~\ref{FIG:8_SVbas} - \ref{FIG:11_SkT6}, the best
agreement with experiment is obtained for the force SkP$^{\delta}$. Indeed, this force gives a reasonable reproduction of the summed strength in $^{24}$Mg
(together with the force SkM*) and the best reproduction of the summed strength in $^{28}$Si. Besides, SkP$^{\delta}$ describes well the peak energy $E_\text{p}$ of the narrow IS0 resonance in $^{24}$Mg and reasonably in $^{28}$Si. So, with this force,  we simultaneously describe the energy of the narrow IS0 peak and integral strength $\sum_{\nu}B_{\nu}(\text{IS0})$. To the best of our knowledge, this is achieved for the first time.

It is also interesting to inspect contributions of IS0 strength from different energy intervals
to the  energy-weighted sum rule EWSR=$2\hbar^2/m \cdot A \langle r^2 \rangle_0$ summarised in Table \ref{tab-8_EWSR}.  It is seen that the experimental strengths exhaust about $70\%$ of the EWSR in $^{24}$Mg and $77\%$ in $^{28}$Si. The forces SkP$^{\delta}$, SkM* and SV-bas give rather similar results. All calculations exhaust close to $100\%$ EWSR in the model space $0-70$ MeV. In $^{24}$Mg, the calculations overestimate the experimental strength at $0-16$ MeV (low-energy IS0 strength and the narrow peak) and underestimate it at $16-25$ MeV (main part of ISGMR). It seems that the Skyrme forces redistribute  too much strength from the main ISGMR to the narrow IS0 peak. A large part of the strength, $25-37\%$, is located above $25$ MeV, i.e. beyond the scope of the iThemba LABS experiment. For $^{28}$Si, the agreement of the experiment and theory is much better. The calculations reproduce rather well a small strength at $0-25$ MeV and its major part at $16-25$ MeV.

\begin{table}[t] 
\caption{Characteristics of low-energy $K^{\pi} = 0^+$ states with excitation energy $E_\nu > 5$ MeV and reduced transition probability $B(\text{IS0}) > 40 ~\rm{fm}^4$ in $^{24}$Mg and $^{28}$Si, calculated within QRPA with the force SkP$^{\delta}$. For each state, we show the excitation energy, the reduced transition probabilities $B(\text{IS0})$ and $B(\text{IS20})$, and the main $2$qp components (contribution to the state norm in $\%$, and structure in terms of Nilsson asymptotic quantum numbers).}
	\begin{center}
\setlength{\arrayrulewidth}{0.5pt}		
\label{tab-8}       
\begin{tabular}{cccccc}
\hline\hline
Nucleus & $E_{\nu}$ &  $B(\text{IS0})$ & $B(\text{IS20})$ &
\multicolumn{2}{c}{main $2$qp components}
\\
\hline
 & (MeV) & (fm$^4$) & (fm$^4$) &\; $\%$ \; & \; $[N,n_z,\Lambda]$ \;  \\
\hline
       & 5.8 &  51  & 56.5 & 49 & pp $[220\uparrow - 211\downarrow]$ \\
        & 12.9 & 75 & 11.7 & 96 & pp $[211\uparrow - 431\uparrow]$ \\
$^{24}$Mg & 13.5 & 77 & 26.5 & 92 & pp $[220\uparrow - 440\uparrow]$ \\
       & 14.3 & 166 & 77.1  & 40 & pp $[211\uparrow - 411\uparrow]$ \\
       & 14.5 & 81 & 40.0 & 48 & pp $[211\uparrow - 411\uparrow]$ \\
\hline
           & 15.3 & 42 & 3.0 & 91 &  pp $[200\uparrow - 431\downarrow]$ \\
           & 16.6 &  59 & 7.1 & 90 & pp $[202\uparrow - 413\downarrow]$ \\
 $^{28}$Si & 16.9 &  91 & 10.5 & 68 &  nn $[200\uparrow - 411\downarrow]$ \\
           & 17.3 & 47 & 4.3  & 39 & nn $[200\uparrow - 420\uparrow]$ \\
           & 17.6 & 62  & 6.0 & 75 & pp $[202\downarrow - 431\uparrow]$ \\
           & 17.8 & 82  & 18.8 & 22 & pp $[200\uparrow - 600\uparrow]$ \\
\hline
\end{tabular}
\end{center}
\end{table}

\subsection{Features of individual IS0 transitions}
\label{s43}

The energy range $9 - 25$ MeV embraces many QRPA $0^+$ states. For example, in the SkP$^{\delta}$
case, we have $631$ states in $^{24}$Mg and $537$ states in $^{28}$Si. It is instructive to consider in more detail  QRPA states which form the narrow IS0 peak and strong IS0 excitations at a lower energy.  In Table \ref{tab-8}, the characteristics of IS0 states in the energy range $5-18$ MeV
with a strength $B(\text{IS0}) > 40$ fm$^4$ are  presented. In $^{24}$Mg, there are two transitions with a large monopole strength below $13$ MeV.
The three transitions at energies between $13.5$ MeV and $14.5$ MeV form the strong narrow IS0 peak shown
in Fig.~\ref{FIG:9_SkP}. Following our calculations, two of these states, at $14.3$ and $14.5$ MeV,
are very collective (composed by many 2qp components with the largest 2qp contributions to the state norm as $40\%$ and $48\%$, respectively), which justifies the collective effect mentioned above.  Most probably just these two states with close energies correspond to the $13.87$ MeV level from the iThemba LABS data (see Table \ref{table:5}). It is difficult to establish one-to-one correspondence between the measured transitions in Table \ref{table:5} and calculated transitions in Table \ref{tab-8} since our Skyrme QRPA calculations neglect some important effects (coupling with complex configurations and shape coexistence) and thus have not enough accuracy for such a detailed comparison. For $^{24}$Mg and $^{28}$Si, there are attempts of one-to-one comparison of the experimentally observed discrete $0^+$-levels using Gogny QRPA \cite{Zabora2021} and Antisymmetrised Molecular Dynamics (AMD) \cite{Chiba_PRC2015} calculations. However, these calculations achieve a reasonable agreement with the data using only an ad-hoc upshift of the IS0 strength distribution by $2$ MeV.

All the states in Table  \ref{tab-8} have large values for $0^+_{\text{gs}} \to 2^+0_{\nu}$ isoscalar quadrupole transitions with $\Delta K =0$. This once more confirms the strong IS0/IS2 coupling in strongly prolate $^{24}$Mg. The main IS0 resonance at $18-25$ MeV (not considered in  Table \ref{tab-8}) is formed
by a large number of mildly collective states with $B(\text{IS0}) < 1$ fm$^4$. The results for strongly oblate $^{28}$Si  generally agree with those for $^{24}$Mg. However, as seen from  Table \ref{tab-8}, in  $^{28}$Si our calculations do not predict monopole states with a high $B(\text{IS0})$ and $B(\text{IS2},K = 0)$ below $15$ MeV.

\subsection{Cluster features of the narrow IS0 peak in $^{24}$Mg}

\begin{figure} 
		\centering
		\includegraphics[width=8cm]{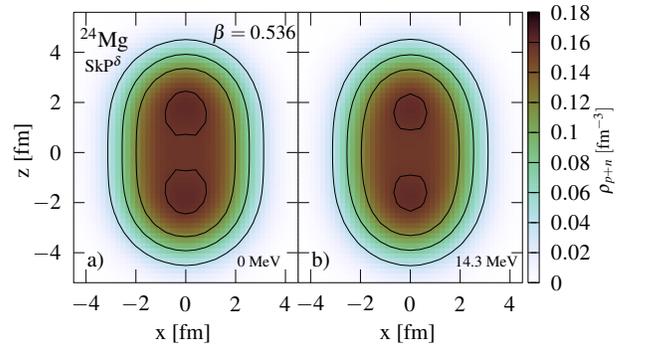}
		\caption{Density profiles in $x-z$ plane, calculated with the force SkP$^{\delta}$ in $^{24}$Mg for the ground state (left) and excited QRPA $K^{\pi} = 0^+$ state at $14.3$ MeV (right).}
		\label{FIG:13_clustering}
\end{figure}

The strong narrow peak at $13.87$ MeV in $^{24}$Mg lies close to the thresholds of cluster configurations $^{12}\text{C}$+$^{12}\text{C}$ and $^{16}\text{O} + 2\alpha$ and, following
the coincidence experiment \cite{kaw2013}, should exhibit cluster features. To check this point, we calculated the density distribution of the QRPA $14.3$ MeV $0^+$ state (assuming correspondence to the observed $13.87$ MeV excitation) and compared it with the density of the ground state. In Fig.~\ref{FIG:13_clustering}, both densities  are exhibited in the $x-z$ plane where $z$ is the symmetry axis. Another relevant collective $0^+$ state at $14.5$ MeV has  a similar density contribution and so is not considered here.

In the ground state, prolate $^{24}$Mg has two spots of the enhanced density in the pole regions. These spots persist, though being  somewhat shrunk, in  the $14.3$ MeV $0^+$ state. The density spots can be considered as precursors of clustering. They can be treated as predecessors of two $^{12}\text{C}$ clusters or, to a lesser extent, of two $\alpha$-particles with $^{16}$O. Note that the plot for $14.3$ MeV state agrees well with the $^{12}\text{C}$+$^{12}\text{C}$ density plots in Ref.~\cite{Chiba_PRC2015} where the cluster features of $^{24}$Mg are analysed within the AMD method.
The similarity of the density plots for the ground and excited states can be explained by a dual nature (mean field + clustering) of the ground state of light nuclei, see the discussion and further citation
in Ref.~\cite{Chiba_PRC2015}. Altogether we see that the strong narrow IS0 peak in $^{24}$Mg might combine both mean-field (IS0/IS2 coupling) and cluster features.  The nucleus $^{28}$Si also demonstrates cluster properties (see e.g. discussions in Refs.  \cite{adsley2021isoscalar,Taniguchi_PRC2009}).
However analysis of these properties is complicated by the shape coexistence in this nucleus and so is skipped here.

\section{Conclusions}
	\label{s5}

The isoscalar monopole strength in the energy interval $9 \leq E_\text{x} \leq 25$ MeV in $^{24}$Mg and $^{28}$Si has been investigated using $\alpha$-particle inelastic scattering with a $196$ MeV beam at scattering angles $\theta_{\text{Lab}} = 0^{\circ}$ and $4^{\circ}$. The K$600$ magnetic spectrometer
and the Separated Sector Cyclotron (SSC) at iThemba LABS were used.
	
The DoS technique was applied in order to extract the IS0 strength distributions. The difference spectrum obtained for each nucleus was converted to the fraction of the IS0 EWSR per bin and compared with MDA analysis results obtained by the TAMU \cite{youngblood2009isoscalar,youngblood2007isoscalar} and RCNP \cite{24Mg_GG_PLB2015,24Mg_GG_PRC2016,28Si_PGG_PRC2016,Zabora2021,kaw2013} groups. Overall, the strength distributions obtained in this study show a reasonable agreement with results from both groups, with exception of the RCNP data
\cite{24Mg_GG_PLB2015,24Mg_GG_PRC2016} for IS0 structures in the energy range $13-18$ MeV in  $^{24}$Mg. At the same time, we find good agreement for this energy range with another $(\alpha, \alpha')$ experiment performed at RCNP \cite{kaw2013}.

The extracted IS0 strength distributions were compared to calculations performed in the framework of QRPA. A representative set of Skyrme forces (SkM*, SV-bas, SkP$^{\delta}$, SkT6 and SV-mas10) with different incompressibility $K_{\infty}$ values
and isoscalar effective masses $m^*_0/m$ was used. Similar to a recent study for Mo isotopes
\cite{E0_Colo_PLB2020}, the present iThemba LABS experimental data for the ISGMR in $^{24}$Mg
and $^{28}$Si are best described by the force SkP$^{\delta}$ with a low incompressibility
$K_{\infty} = 202$ MeV. This force allows the reproduction of both i) the energy of the narrow IS0 peaks
at $13.8$ MeV in $^{24}$Mg and $18$ MeV in $^{28}$Si  and ii) the integral IS0 strengths
$\sum B(\text{IS0})$.

\begin{figure*} 
		\centering
		\includegraphics[height=10cm,width=14cm]{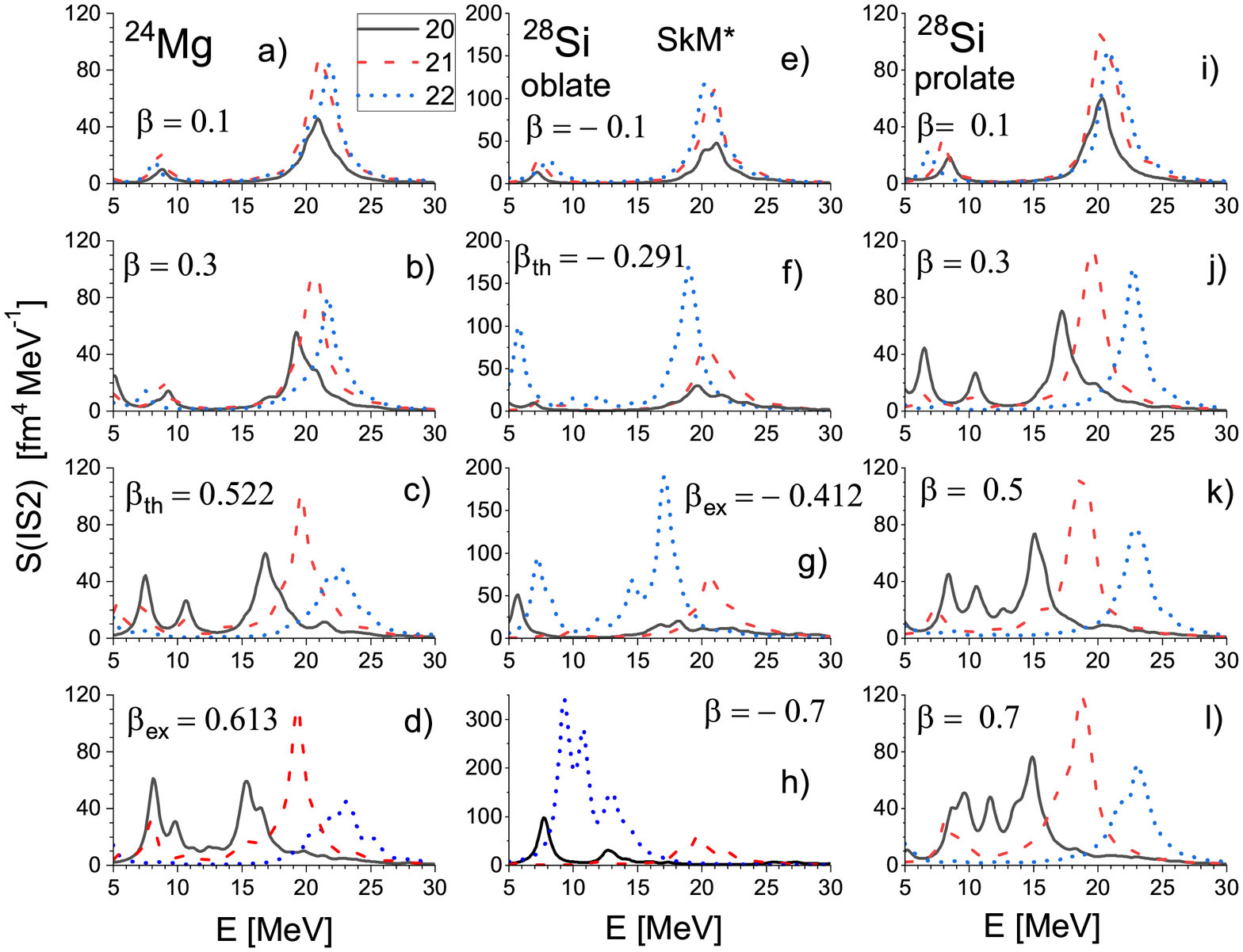}
		\caption{ISGQR branches $LK =2 0$ (solid black line), $21$ (dashed red line) and $22$ (dotted blue line) in $^{24}$Mg  and $^{28}$Si, calculated with the force SkM* for different quadrupole deformations. For $^{28}$Si, both oblate (middle panels) and  prolate (right panels) deformations are considered.}
		\label{FIG:14}
	\end{figure*}

In the theoretical analysis, the main attention was paid to the narrow IS0 peak constituting
a convenient and sensitive test case to determine the nuclear incompressibility in deformed nuclei.
The comparison of IS0 and IS2 strength distributions justifies that the narrow IS0 peak appears
due to the deformation-induced coupling between the ISGMR and the $K = 0$ branch of the ISGQR. The important role of a large collective downshift of the IS0 strength (due to the isoscalar residual interaction) in the production of the narrow IS0 peak is demonstrated. In $^{24}$Mg, the narrow IS0 peak  lies close to the $^{12}\text{C}$+$^{12}\text{C}$ threshold \cite{kaw2013} and so, in accordance with our analysis and observations \cite{kaw2013},
can demonstrate cluster features.

In connection with the strong IS0/IS2 coupling effect, we also performed a detailed
analysis of the ISGQR in $^{24}$Mg and $^{28}$Si. An unusual deformation splitting in oblate $^{28}$Si was found. The calculations show that the quadrupole $K = 0$ branch is strongly downshifted  in prolate $^{24}$Mg and slightly downshifted in oblate
$^{28}$Si. This explains why the narrow IS0 peak is well separated from the main ISGMR in $^{24}$Mg but superimposed on the broad structure of the main ISGMR in $^{28}$Si.

Our analysis demonstrates that the ISGMR in deformed nuclei must be described simultaneously with the ISGQR. Only then can such a description be considered as relevant and consistent. In our calculations, the Skyrme force SkP$^{\delta}$ gives the best results for a simultaneous reproduction of ISGMR and ISGQR. However, even this force has shortcomings for the description of the ISGQR. There is a need for new Skyrme parameterisations with $K_{\infty} < 217$ MeV and $m^*_0/m \approx 0.9$.

\section*{ACKNOWLEDGEMENTS}
	
The authors thank the Accelerator Group at iThemba
LABS for the high-quality dispersion-matched beam provided for this experiment. A.B. acknowledges
financial support through iThemba LABS, NRF South Africa. R.N. acknowledges support from the NRF through Grant No. $85509$. The authors thank Prof P.-G. Reinhard for the code SKYAX. This work was supported by the Deutsche Forschungsgemeinschaft under contract SFB $1245$ (Project ID No. $79384907$) and NRF-JINR grant JINR200401510986. V.O.N. and J.K. acknowledge the  Votruba - Blokhintsev (Czech Republic - BLTP JINR) grant. J.K. appreciates the support by a grant of the Czech Science Agency, Project No. 19-14048S. V.O.N. and P.v.N.-C. appreciate the support by a Heisenberg-Landau grant (Germany - BLTP JINR). A.R. acknowledges support by the Slovak Research and Development Agency under Contract No. APVV-20-0532 and by the Slovak grant agency VEGA (Contract No. 2/0067/21). P. A. acknowledges support from the Claude Leon Foundation in the form of a postdoctoral fellowship.
	
\appendix

\section{ISGQR in $^{24}$Mg and $^{28}$Si}
\label{s6}

As shown in the main text, the $K = 0$ branch of the ISGQR determines the energy of the narrow IS0 peak. The energy of this branch is governed by the deformation splitting of the ISGQR into $K = 0, 1$ and $2$ components. So, it is worthwhile to investigate the deformation splitting of the ISGQR in $^{24}$Mg and $^{28}$Si in more detail. Figure \ref{FIG:14} shows the strength distributions of the branches for different axial quadrupole deformations. The calculations are performed for the parametrisation SkM* as an example but other forces lead to the same qualitative conclusions.

In the plots a)-d), the ISGQR strength distribution in prolate $^{24}$Mg is exhibited at $\beta = 0.1$, $0.3$, $0.522$ (calculated equilibrium value) and  0.613 (experimental value). We see a splitting  picture typical for prolate nuclei: the $K = 0$ branch is downshifted from $21$ MeV ($\beta = 0.1$) to $16$ MeV ($\beta = 0.613$), the $K = 1$ branch shows less downshift from $21$ to $19$ MeV, and the $K = 2$ branch is upshifted from $22$ to $23$ MeV. The energy of the $K = 0$ branch is most sensitive to the deformation. However, even at $\beta = 0.613$, this branch does not approach an excitation energy $13-14$ MeV where experiment gives the narrow monopole peak. So, to describe the experimental data we need a force like SkP$^{\delta}$ with a smaller incompressibility.

       \begin{table} [t]
		\caption{EWSR exhaustion for the $LK = 20$, $21$ and $22$ branches of the ISGQR, calculated with the force SkM*.}
			\begin{center}
   \vspace{0.2cm}
		\label{tab-9}       
		\setlength{\arrayrulewidth}{0.5pt}
		\setlength{\tabcolsep}{0cm}
		\renewcommand{\arraystretch}{2}	
		\scalebox{0.94}{$\begin{tabular}{ccccc}
			\hline\hline
                    &  \multicolumn{2}{c}{ \;$^{24}$Mg, $\beta = 0.522$ \;} & \multicolumn{2}{c}{\;$^{28}$Si, $\beta =  -0.291$ \;} \\
        \hline
                    & \; ($\rm{fm}^4$ MeV) &\: (\%) \; & \; ($\rm{fm}^4$ MeV) &\: (\%) \;  \\
			\hline
            \; 20 \; & 4468 & 26.5  & 3650 & 17.5 \\
            \; 21 \;&  7396 & 42.0  & 6998 & 33.7  \\
            \; 22 \; & 5534 & 31.5  & 10144  & 48.8 \\
            \hline
			\end{tabular}$}
		\end{center}
	\end{table}

In the plots e)-h), the dependence of the ISGQR in $^{28}$Si on an {\it oblate} quadrupole deformation is shown. The deformations are  $\beta = -~0.1$,
$-~0.291$ (calculated equilibrium values),  $-~0.412$ (experimental value) and $-~0.7$. In the oblate case, the deformation splitting demonstrates some surprising features. The $K = 0$ branch is not upshifted, as would be expected from analytic estimates \cite{E0_Kvasil_PRC2016}, but downshifted from $21$ MeV to $\sim 13$ MeV. The $K = 0$
strength is redistributed to a lower-energy region and significantly suppressed (see also Table \ref{tab-9}). The $K = 1$ branch keeps almost the same energy. The $K = 2$ branch is strongly downshifted from $21$ MeV to $8-15$ MeV and its strength is significantly enhanced (see also Table \ref{tab-9}). At the calculated ($\beta_{\text{th}} = -~0.291$) and experimental ($\beta_{\text{ex}} = -~0.412$) values of the quadrupole deformation,
the $K = 0$ peak approaches the energy $17-18$ MeV of the narrow monopole peak. So the force SkM* can in principle describe the narrow IS0 peak in $^{28}$Si once we take a proper value of oblate quadrupole deformation.

In the plots i)-l), the dependence of ISGQR in $^{28}$Si on a {\it prolate} quadrupole
deformation is inspected. This case is interesting since some studies predict at $9-13$ MeV in $^{28}$Si
superdeformed rotational bands based on prolate $0^+$ band heads \cite{adsley2017alpha,Taniguchi_PRC2009,Taniguchi_PLB2020,Afan+Ray}. As seen from Fig.~\ref{FIG:7ensurf}, the total energy of $^{28}$Si indeed has a local prolate minimum. The plots i)-l) show that the deformation splitting in prolate $^{28}$Si would be rather similar to that in prolate  $^{24}$Mg. The $K = 0$ branch in prolate $^{28}$Si indeed covers the relevant energy interval $8-16$ MeV. Our calculations, therefore, do not contradict the suggestion \cite{adsley2017alpha,Taniguchi_PRC2009,Taniguchi_PLB2020,Afan+Ray} that $^{28}$Si can possess
highly prolate states at excitation energies $9-13$ MeV. Moreover, at $\beta = 0.3$, the $K = 0$ strength covers the energy interval $16-19$ MeV  where the experimental narrow IS0 peak is located. So, our SkM*
calculations do not contradict that $^{28}$Si, being oblate in the ground state, can have a prolate shape for $0^+$ excitations.

Altogether, the present analysis allows for the conclusion that the description of the experimental narrow IS0 peaks: i) in the case of $^{24}$Mg, needs a Skyrme force like SkP$^{\delta}$ ($K_{\infty} = 202$ MeV) with a low nuclear incompressibility and ii) in the case of $^{28}$Si, can be achieved  with both SkP$^{\delta}$ and  SkM* ($K_{\infty} = 217$ MeV) forces once we allow for oblate and prolate quadrupole deformation.

\end{document}